\documentclass[aps, prb, unsortedaddress,floatfix, nofootinbib, twocolumn, superscriptaddress,floatfix, longbibliography,colorlinks, breaklinks,
linkcolor=NavyBlue,
citecolor=NavyBlue,
urlcolor=NavyBlue,10pt]{revtex4-2}
\usepackage{graphicx, color, soul}
\usepackage{amsmath,amssymb,bm}
\usepackage[usenames,dvipsnames]{xcolor}
\usepackage{ulem}
\usepackage{multirow}
\usepackage{color}
\usepackage{bm}
\usepackage[mathscr]{eucal}
\usepackage{physics}
\usepackage[version=4]{mhchem}
\usepackage{multirow}
\usepackage{subfiles}
\usepackage[utf8]{inputenc}
\usepackage{xltabular}
\usepackage{tabularx}
\usepackage{tikz}
\usepackage{makecell}
\newcommand{\eq}[1]{\begin{align}#1\end{align}}
\usepackage{longtable}
\newcolumntype{M}[1]{>{\centering\arraybackslash}m{#1}}
\usepackage{float}
\usepackage{algorithm}
\usepackage{algpseudocode}
\usepackage{orcidlink}
\usepackage{hyperref}
\usepackage{hypcap}
\usepackage{tcolorbox}
\tcbuselibrary{breakable}
\newcommand{\rev}[1]{\textcolor{black}{#1}}
\newtcolorbox{scopebox}{
    breakable,
    colback=gray!5,
    colframe=black,
    boxrule=0.8pt,
    arc=2mm,
    left=2mm,
    right=2mm,
    top=2mm,
    bottom=2mm,
    title=\textbf{Scope of the Work},
    fonttitle=\bfseries
}

\begin{document}

\title{Meta-learning of Gibbs states for many-body Hamiltonians with applications to Quantum Boltzmann Machines}

\author{Ruchira V Bhat \orcidlink{0000-0002-3474-2286}}
\affiliation{Quantum Lab, Fujitsu Research of India}
\email{Ruchira.Bhat@fujitsu.com}

\author{Rahul Bhowmick \orcidlink{0000-0002-5893-0847}}
\affiliation{Quantum Lab, Fujitsu Research of India}
\author{Avinash Singh \orcidlink{0009-0009-9116-2884}}
\affiliation{Quantum Lab, Fujitsu Research of India}

\author{Krishna Kumar Sabapathy \orcidlink{0000-0003-3107-6844}}
\affiliation{Quantum Lab, Fujitsu Research of India}

\date{\today}

\begin{abstract}
The preparation of quantum Gibbs states is a fundamental challenge in quantum computing, essential for applications ranging from modeling open quantum systems to quantum machine learning.
Building on the Meta-Variational Quantum Eigensolver framework proposed by Cervera-Lierta et al.\cite{meta-vqe} and a problem-driven ansatz design, we introduce two meta-learning algorithms : Meta-Variational Quantum Thermalizer (Meta-VQT) and Neural Network Meta-VQT (NN-Meta VQT) for efficient thermal state preparation of parametrized Hamiltonians on Noisy Intermediate-Scale Quantum (NISQ) devices.
\rev{Meta-VQT utilizes a fully quantum ansatz, while NN Meta-VQT integrates a quantum--classical hybrid architecture. Both use collective optimization over a finite set of training parameter values to prepare approximate Gibbs states at held-out parameter values within the sampled parameter domain.}
\rev{We validate our methods on Transverse Field Ising Models (TFIM) with up to eight qubits and on a two qubit Heisenberg model with all field terms. Within these finite-size benchmarks, the trained models accurately prepare thermal states at test points inside the training interval. For the six and eight qubit cases of TFIM, meta-learned parameters when combined with appropriately designed ansatz serve as warm-start initializations, significantly outperforming random initializations in the optimization tasks.}
Furthermore, a 3 qubit Kitaev ring example showcases our algorithm’s effectiveness across finite-temperature crossover regimes.
Finally, we apply our algorithms to train a Quantum Boltzmann Machine (QBM) on a 2 qubit Heisenberg model with all field terms, achieving enhanced training efficiency, improved Gibbs state accuracy, and a 30-fold runtime speedup over existing techniques such as variational quantum imaginary time (VarQITE)-based QBM highlighting the practicality of meta-algorithm-based QBMs.
\end{abstract}

\maketitle

\section{Introduction}
Preparation of quantum Gibbs states is of fundamental importance as they provide statistical description of quantum systems at finite temperature, which is the natural setting for almost all real-world physical systems.
Beyond their foundational role in statistical mechanics such as the study of open quantum systems \cite{poulin2009sampling}, Gibbs states have emerged as valuable resources in quantum technologies, underpinning applications in areas such as quantum simulation \cite{childs2018toward}, quantum machine learning \cite{bound_based_qbm1, biamonte2017quantum,amin_qbm,Zoufal_circuit_training,bound_based_qbm2,state_based_1,state_based_2,state_based_3}, optimization \cite{combinatorial_optimization}, quantum metrology \cite{quantum_metrology} and quantum cryptography \cite{quantum_key_dis}. Their broad utility, coupled with the intrinsic difficulty of preparing them efficiently \cite{aharonov2013guest}, has motivated an extensive line of research aimed at developing algorithms and protocols for Gibbs state generation using classical and quantum computational techniques \cite{terhal2000problem, poulin2009sampling,temme2011quantum,kastoryano2016quantum,brandao2019finite,eftqc1,eftqc2,ftqc1,wiebe2014quantum,yung2012quantum,kaplan2017ground,riera2012thermalization}.

\begin{figure*}[!tbh]
    \centering
    \includegraphics[scale=0.26]{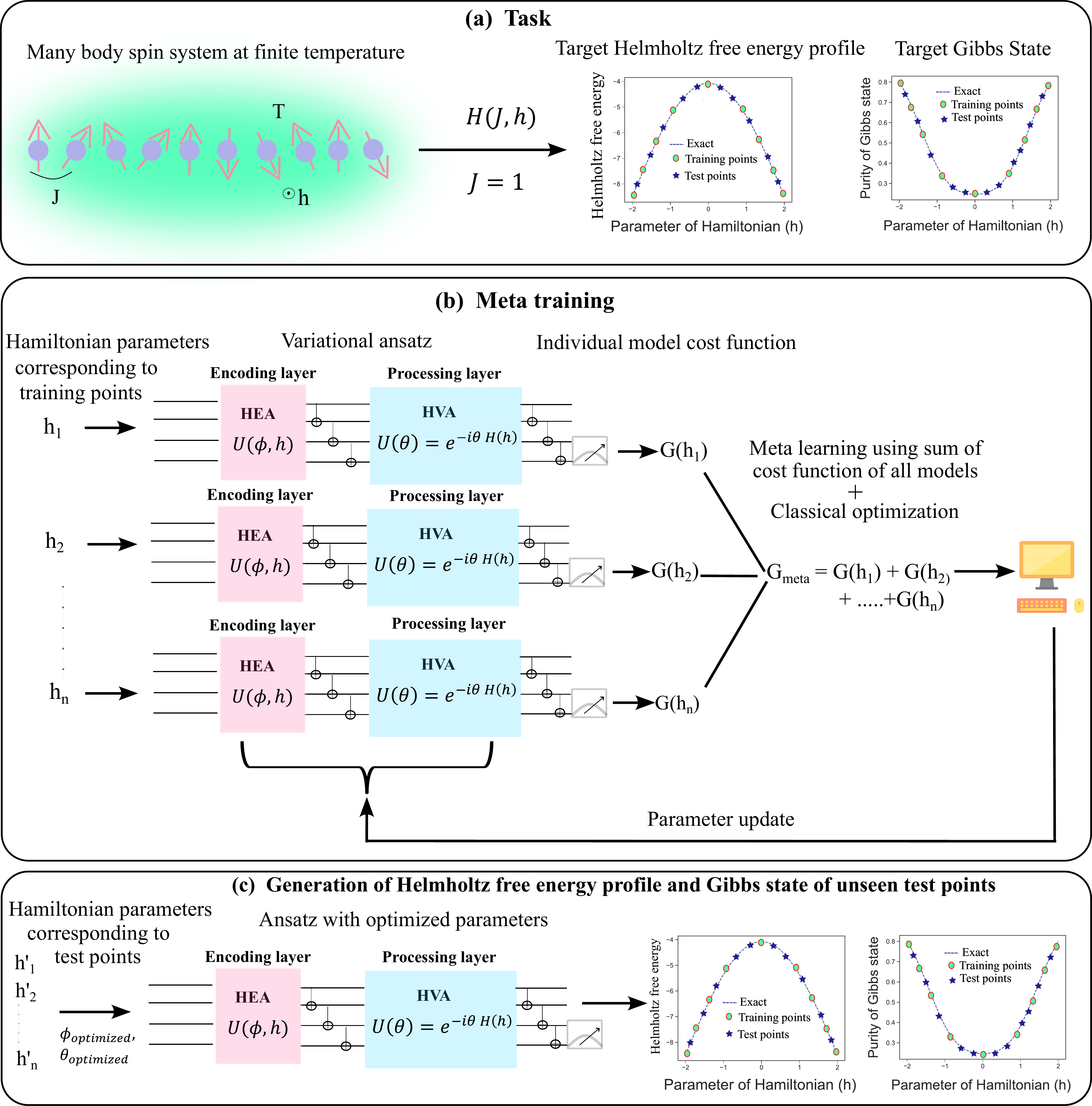}
    \caption{\textbf{Overview of meta-algorithm for Gibbs state preparation} :(a) Depicts the core task addressed by our meta-algorithm: preparing thermal states for a parametrized many-body spin Hamiltonian at finite temperature. Shown are illustrative profiles of the Helmholtz free energy and the purity of the corresponding Gibbs state as functions of a Hamiltonian parameter. The parameter space is split into training (green circles) and test (blue star) sets. (b) Illustrates the training process, where parameters from the training set are encoded into a variational ansatz composed of a hardware-efficient and a Hamiltonian variational ansatz, separated by a linear arrangement of CNOT gates. The global cost, evaluated across all training points, is minimized using a classical optimizer. (c) Shows inference, where test parameters are input into the trained ansatz, initialized with the optimized parameters from (b), to generate Gibbs states. These are then used to reconstruct the Helmholtz free energy profile.}
    \label{meta_algorithm_overview}
\end{figure*}

While many of these existing quantum computing approaches are designed with fault-tolerant quantum hardware in mind, near-term devices are constrained by noise and limited circuit depth.
Variational Quantum Algorithms (VQAs) that can be implemented efficiently on noisy intermediate-scale quantum (NISQ) devices, present an alternative to these methods.
VQAs represent a class of hybrid algorithms that combine quantum and classical resources. The quantum device evaluates a loss function determined by the circuit parameters, and a classical optimizer updates these parameters iteratively.

Several variational techniques have also been proposed for Gibbs state preparation on near-term quantum hardware. These include Variational Quantum Thermalizers (VQTs) which are the finite-temperature extensions of the variational quantum eigensolver (VQE) \cite{verdon_vqt,HVA1,HVA2,HVA3}, methods that utilize variational quantum imaginary time evolution (VarQITE) \cite{mcardle2019variational, yuan2019theory, Zoufal_circuit_training}, and truncated Taylor series expansions of the Helmholtz free energy as a cost function \cite{wang2005variational}.
Additionally, dynamic parameterized quantum circuits (PQCs) that mitigate barren plateaus \cite{baren_plateau_free_dynamic_pqc}, and non-unitary multi-qubit operations \cite{multiqubit_nonunitary} are proposed to enhance algorithm performance for Gibbs state preparation.
A significant drawback of these algorithms is that they must be executed separately for each Hamiltonian instance, making them computationally expensive when preparing Gibbs states for a family of parameterized Hamiltonians—for example, during the training of QBMs \cite{amin_qbm, Zoufal_circuit_training}, or when exploring thermal properties across varying coupling strengths to construct many-body phase diagrams \cite{many_body_phase_1,sachdev2011quantum}.
Another challenge is to select an ansatz with sufficient expressivity to efficiently prepare the Gibbs state for specific Hamiltonian parameters and temperatures, especially when the Hamiltonian undergoes finite-temperature crossovers as a function of those parameters \cite{simulation_quantum_critical}.

\rev{In this work, we address the repeated optimization and ansatz design challenges by introducing two algorithms for preparing Gibbs states of parameterized Hamiltonians:}
\begin{itemize}
    \item Meta-Variational Quantum Thermalizer (Meta-VQT), and
    \item Neural Network Meta-Variational Quantum Thermalizer (NN Meta-VQT).
\end{itemize}
Both methods are inspired by Meta-VQE \cite{meta-vqe}, NN Meta-VQE \cite{nn-meta-vqe}, and related collective optimization approaches \cite{coll_opt_vqe1,coll_opt_vqe2}, which learn ground state energy profiles at zero temperature. \rev{Our central methodological contribution is to extend this collective optimization (meta-learning) principle to mixed finite-temperature state preparation. Meta-VQT learns a parameter-conditioned quantum circuit, whereas NN Meta-VQT learns a classical map from Hamiltonian parameters to quantum circuit parameters. Unlike supervised state learning schemes, neither method receives exact Gibbs states, density matrices, matrix elements, or thermodynamic observables as training labels.
For a finite set of Hamiltonian parameters, both methods construct a global cost function by summing the Helmholtz free energy objectives associated with the individual parameter instances. Collective optimization of this objective produces a single parameter-conditioned state preparation strategy, rather than requiring an independent variational optimization for each Hamiltonian.}
\rev{The resulting model can therefore be evaluated at both training and unseen test (held-out) parameter values within the training interval. This differs from approaches such as that of \textcite{meta-vqe}, which learn a scalar ground state energy landscape rather than a state preparation strategy. Because our approach explicitly prepares an approximate Gibbs state, its performance can be assessed beyond the free energy objective. We therefore evaluate the prepared states using state fidelity, trace distance and purity, in addition to the Helmholtz free energy. Moreover, once the state has been prepared, expectation values of other observables and, in principle, additional thermodynamic quantities can be evaluated without retraining the model. The overall workflow for training and evaluating the resulting Gibbs state preparation model is illustrated in Fig.~\ref{meta_algorithm_overview}.}

We investigate the methods using the following numerical examples:
\begin{itemize}
    \item a single parameter transverse field Ising model (TFIM) with system sizes from two to eight qubits;
    \item a three qubit, single parameter Kitaev ring whose sampled parameter interval contains a finite-temperature crossover \cite{kiatev_ring2}; and
    \item a two qubit Heisenberg Hamiltonian with two parameters, all field terms, and three commuting blocks, including its use in a QBM workflow.
\end{itemize}
In these numerical studies, we use same number of ancilla as the system qubits. The resulting total qubit counts are comparable to the finite-size variational Gibbs state studies summarized in Table~\ref{gibbs_state_exisiting_lit}. \rev{This choice provides a purification space sufficient for the tested mixed states.
The numerical evidence has two significant parts. First, after collective training, Meta-VQT and NN Meta-VQT prepare states close to the exact Gibbs states at held-out (unseen) Hamiltonian parameter values within the training interval. Second, when direct evaluation with the trained model is not sufficiently accurate, the meta-trained parameters provide a warm-start for a conventional single instance VQT. For the tested six and eight qubit TFIM examples, this initialization outperforms the two randomly initialized variational baselines considered in Sec.~\ref{better_initialization}.}

Finally, we demonstrate the utility of these algorithms by improving the training of a variational QBM with a simple example: a 2 qubit double parameterized Heisenberg Hamiltonian with external fields.
Upon comparison of our results with an existing technique of VarQITE-based variational QBM, we observe that our collective optimization algorithms offer several advantages. Specifically,
\begin{itemize}
    \item Enhance the training speed and  reduce the complexity by requiring fewer QPU calls,
    \item Increase the training efficiency by more accurate Gibbs state preparation, as the system size increases and the Hamiltonian and input data distributions become more complex.
\end{itemize}
For relatively simple Hamiltonians with fewer non-commuting terms, our approach has performance comparable to existing methods. However, as the complexity of the Hamiltonian grows, our proposed protocol have the potential to offer efficient performance, highlighting its broader applicability and robustness for practical quantum machine learning tasks.

\begin{scopebox}
\rev{\textit{Objective :} We develop two meta-learning approaches, Meta-VQT and NN Meta-VQT, for finite-temperature quantum state preparation. Both methods learn a variational thermal state preparation strategy that adapts to different Hamiltonian parameter values. Training is performed collectively over a finite set of Hamiltonian instances by encoding the parameter values and minimizing the corresponding Helmholtz free energy objectives, without requiring exact Gibbs states, density matrices, or thermodynamic observables as training labels.}

\rev{\textit{Novelty : } We extend collective optimization from ground state energy profile learning to mixed finite-temperature state preparation and combine it with thermal state benchmarks and an end-to-end QBM application.}

\rev{\textit{Numerical Examples and Main Results : } Two to eight qubit TFIM test Gibbs state preparation at held-out parameters and warm-start performance; a three qubit Kitaev ring tests Gibbs state preparation over a finite-temperature crossover regime; and a two qubit Heisenberg QBM demonstrates an end-to-end application with a benchmark specific thirty-fold runtime reduction.}

\rev{\textit{Limitations :}  The present results are demonstrated for finite-size Hamiltonian models, and the extent to which the learned Gibbs state preparation strategies generalize to larger system sizes and parameters outside the training interval remains to be established. Asymptotic scalability and quantum advantage are therefore not claimed at this stage. Addressing these questions will require systematic studies across larger systems, broader Hamiltonian families and parameter domains, as well as implementations on relevant quantum hardware.}
\end{scopebox}

\section{Results at a glance}\label{results_at_glance}
\rev{The principal numerical findings are summarized below to establish the scope of the evidence before presenting the technical background and methods. Detailed setups, evaluation procedures, and comparisons are provided in Sec.~\ref{results meta and nn meta} and the Appendices.}
\begin{itemize}
\item \rev{\textit{Ansatz Design:} Two composite ansatz are considered: (i) Meta-VQT : an ansatz combining hardware compatible parameter encoding with Hamiltonian informed layers, and (ii) NN Meta-VQT : an ansatz incorporating classical neural network based parameter encoding with hardware efficient and Hamiltonian informed layers. Both are designed to represent the thermal states of the target Hamiltonians. Their performance on the finite-size benchmarks motivates future development of adaptive, symmetry preserving, and lower depth variants for larger systems.}

\item \rev{\textit{Held-out Parameters:} Meta-VQT and NN Meta-VQT prepare accurate Gibbs states for different many-body Hamiltonians at held-out parameter values within the training interval for systems of upto 4 qubits (for TFIM).}

\item \rev{\textit{Warm-Starts:} For the larger six and eight qubit TFIM instances, meta-trained parameters provide improved initializations for subsequent VQT optimization compared with the two randomly initialized baselines considered.}

\item \rev{\textit{Finite-Temperature Crossover:} On a three qubit Kitaev ring, Meta-VQT prepares Gibbs states at held-out parameter values spanning a finite-temperature crossover regime, demonstrating transfer of the learned state preparation strategy across the crossover region.}

\item \rev{\textit{Application:} In the simulated two qubit Heisenberg QBM benchmark, NN Meta-VQT avoids repeated inner loop Gibbs state preparation and achieves an approximately thirty-fold reduction in runtime relative to the tested VarQITE-based implementation.}

\end{itemize}

\section{Background}
In this section, we briefly review one of the most widely studied NISQ compatible approaches for the preparation of Gibbs states: the VQT introduced by Verdon et al.\cite{verdon_vqt}, which forms the foundation of our proposed meta-algorithms.
The VarQITE framework developed by Zoufal et al.  \cite{Zoufal_circuit_training}, which we use as a baseline for comparative analysis throughout this work is explained briefly in Appendix \ref{varqite}.
\subsection{Variational Quantum Thermalizer (VQT)} \label{VQT section}
Consider a n-qubit Hamiltonian $H(\vec{h})$ at temperature $T$.
Then the Gibbs state of this parametrized Hamiltonian is given as,
\eq{\label{Gibbs state}
\rho(\vec{h}) = \frac{e^{-\beta H(\vec{h})}}{\sum e^{-\beta H(\vec{h})}},}
where $\beta = 1/k_{B}T$ with $T$ being the temperature, $k_{B}$ is the Boltzmann constant and $\sum e^{-\beta H(\vec{h})}$ is the partition function.
For convenience, we assume that the Boltzmann constant is equal to $1$ throughout the paper.

\begin{figure*}[!tbh]
    \centering
    \includegraphics[scale=0.54]{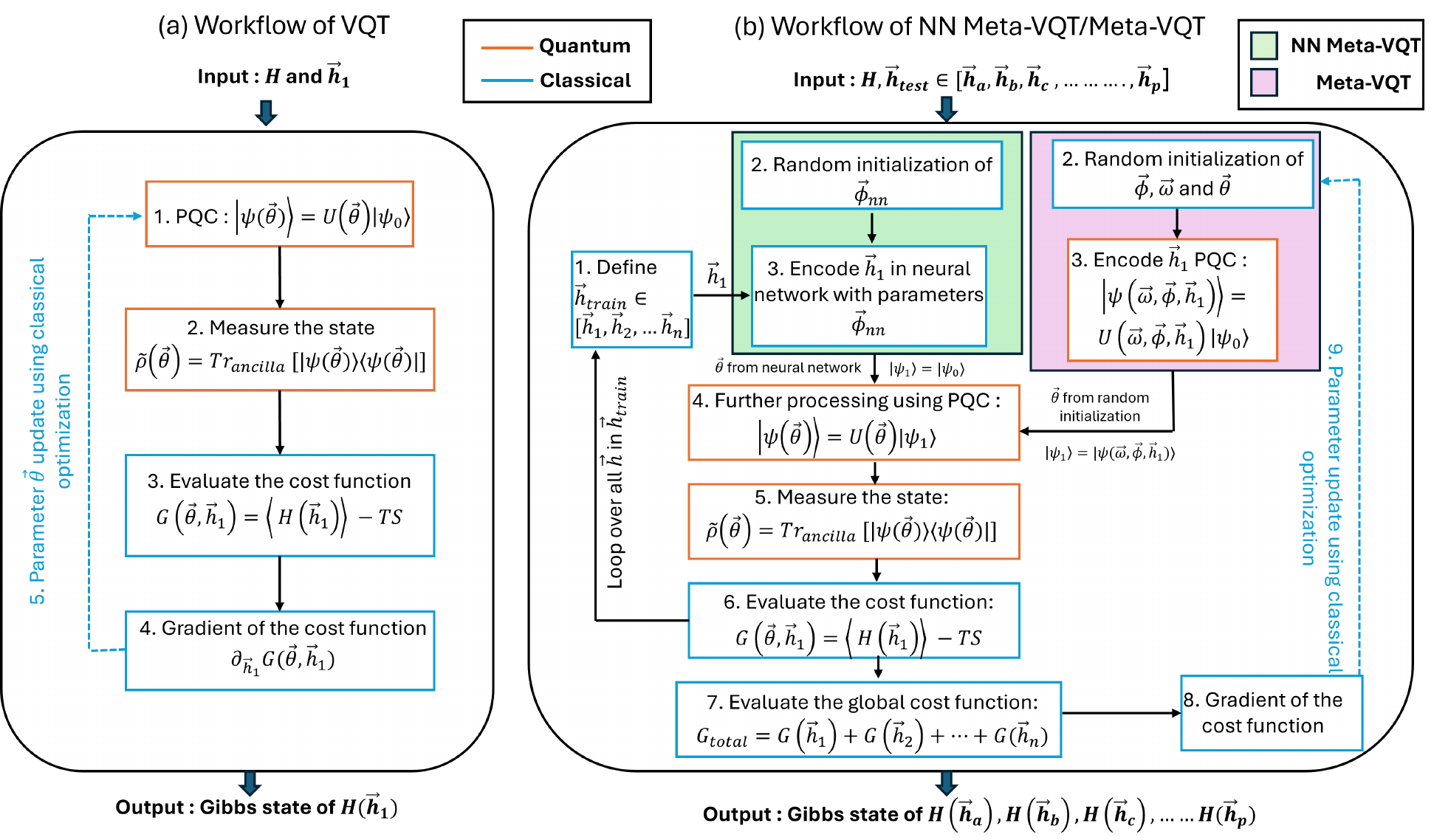}    \caption{\textbf{Workflow of VQT, Meta-VQT and NN Meta-VQT} : The workflow of VQT, where the PQC is trained for approximating the Gibbs state corresponding to a particular parameter value, $\vec{h}_{1}$, of the Hamiltonian is shown in (a). For a different parameter, this workflow needs to be run again. The workflow of Meta-VQT and NN Meta-VQT is shown in (b). Unlike (a) here a set of Hamiltonian parameters $\vec{h}_{train}$ (Step 1), are encoded in the PQC (Step 3, pink colored box) or classical neural network (Step 3, green colored box), and a global loss function is evaluated (Step 7) that does collective optimization over these set of parameters. \rev{After training, the PQC is evaluated at test values $\Vec{h}_{test} \notin \Vec{h}_{train}$. Our primary tests use omitted values within the training interval; performance outside that interval is assessed separately in Appendix~\ref{results_outsidetraining}.} In both (a) and (b), the blue boxes denote modules implemented on a classical computer, while the orange boxes indicate modules that become classically intractable as system size increases and are therefore intended for implementation on a quantum computer.}
    \label{vqt_meta_workflow}
\end{figure*}

Given a PQC represented by unitaries $U(\Vec{\theta})$ with $m$ trainable angles, $\Vec{\theta} \in \theta_{1}, \theta_{2},...\theta_{m}$ acting on $\text{n} + \text{n}_{\text{ancilla}}$ qubits,
the goal of VQT is to train the PQC to approximate the Gibbs state of this Hamiltonian as,
$\rho(\vec{h}) \approx \tilde{\rho} (\Vec{\theta}) = \text{Tr}_{\text{ancilla}} \left(\ket{\psi(\Vec{\theta})}\bra{\psi(\Vec{\theta})}\right)$,
where $\ket{\psi(\Vec{\theta})} = U(\Vec{\theta})\ket{\psi_{0}}$ is the state generated by the PQC from the initial state $\ket{\psi_{0}}$.
The exact loss function to be minimized using a classical computer to train this PQC is the Helmholtz free energy given by,
\eq{ \label{Gibbs free energy}
G(\Vec{\theta},\Vec{h}) = \langle H(\Vec{h})\rangle - T S,
}
where $ \langle H(\Vec{h})\rangle = \text{Tr}[H(\Vec{h})\Tilde{\rho}(\Vec{\theta})] $ is the expectation value of the Hamiltonian, $S = -\text{Tr}[\Tilde{\rho}(\Vec{\theta})\log \Tilde{\rho}(\Vec{\theta})]$ is the Von-Neumann entropy and $T=1/\beta$ is the temperature of the system.
The minimization of $G(\Vec{\theta},\Vec{h})$ ensures that the system reaches the corresponding Gibbs state which is the state of minimal energy at finite-temperature.
The schematic workflow of VQT is shown in Fig.~\ref{vqt_meta_workflow}(a).
This is in analogy with the VQE where the expectation value of the Hamiltonian is minimized to find the ground state energy.
In the limit $T \rightarrow 0$, Eq.~\eqref{Gibbs free energy} reduces to minimization of the expectation value of the Hamiltonian, which is the loss function in VQE.
However, unlike VQE, the mixed nature of the Gibbs state necessitates the use of ancilla qubits $(1~\leq~\text{n}_{\text{ancilla}}~\leq~\text{n})$ to prepare the approximate Gibbs state of given Hamiltonian.

\rev{The evaluation of the Helmholtz free energy, particularly its von Neumann entropy contribution, becomes increasingly challenging as the system size grows. Several approaches have therefore been proposed to approximate or efficiently estimate the free energy using variational, hybrid quantum-classical, or classical methods. A brief discussion of representative approaches is provided in Appendix~\ref{Gibbs free energy evaluation}. Importantly, efficient free energy evaluation is not the primary methodological focus of this work. Our main contribution is the collective optimization of a parameter-conditioned Gibbs state preparation strategy across multiple Hamiltonian instances. The free energy estimation techniques discussed in the Appendix~\ref{Gibbs free energy evaluation} can, in principle, be incorporated into our framework as alternative implementations of the underlying free energy objective, and are therefore complementary to, rather than specific to, our collective optimization approach.}

\begin{figure*}[!tbh]
    \centering
    \includegraphics[scale=0.43]{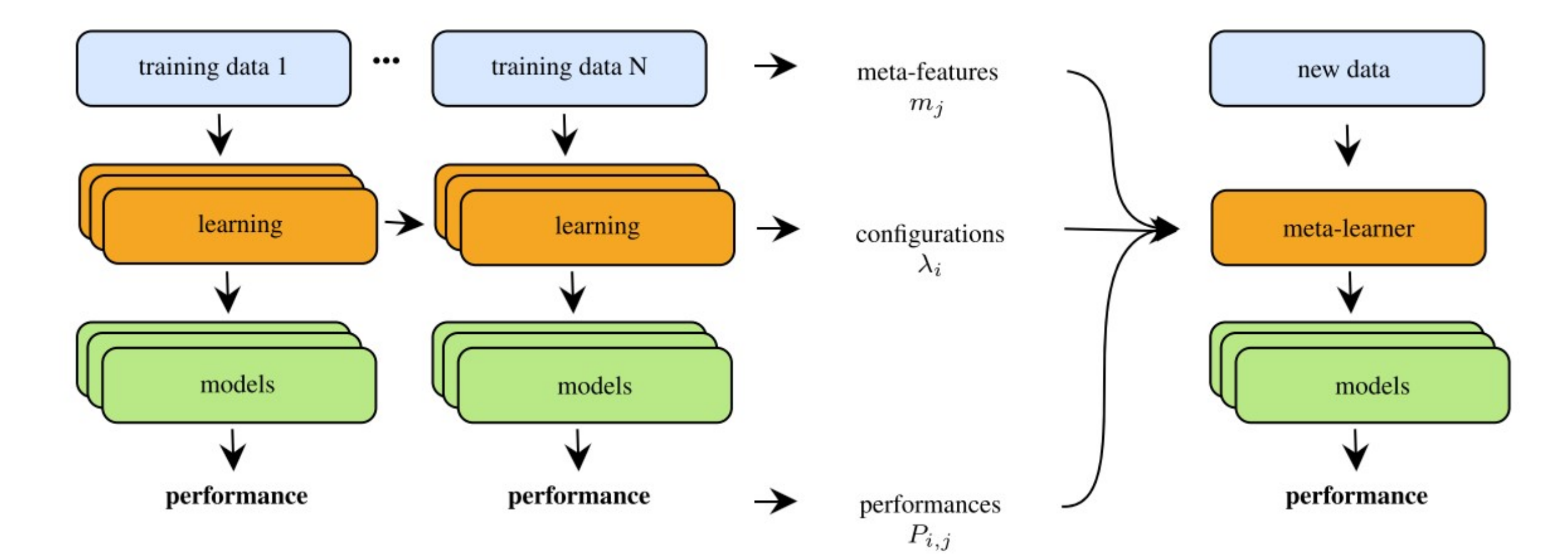}
    \caption{\textbf{Schematic workflow of meta-learning algorithms :} In the workflow presented here, different types of training data denoted by training data 1,..,training data N  are associated with distinct learning algorithms and models, each defined by a separate loss function. The meta-learning approach optimizes a combined loss function based on $P_{i,j}$ and shared feature representations and configurations, $m_{j}~\text{and}~\lambda_{i}$, enabling the resulting meta-learner model to perform effectively on unseen new data that was not included in the training dataset. This schematic workflow has been adapted from Bahri et al.\cite{meta_learning_workflow}.}
    \label{schematics meta}
\end{figure*}
\begin{figure*}[!tbh]
    \centering
    \includegraphics[scale=0.21]{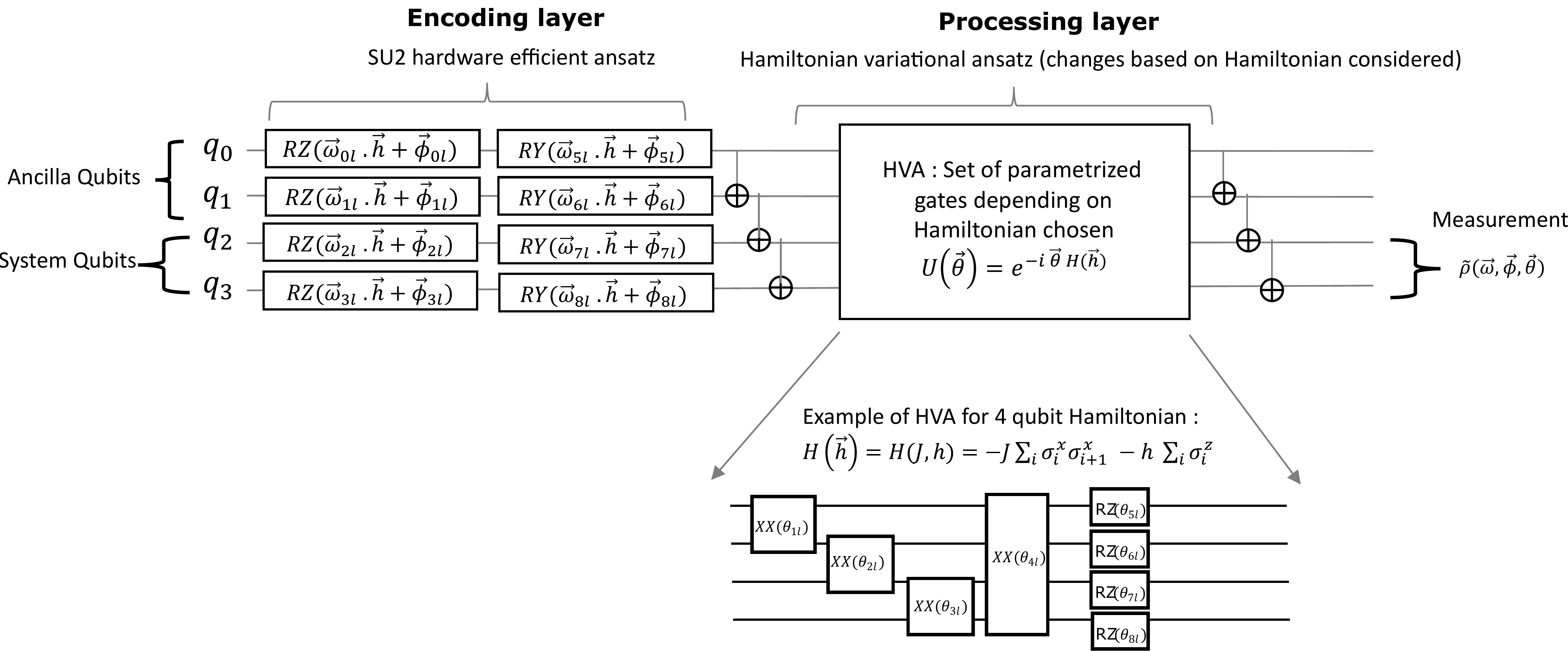}
    \caption{\textbf{Meta-VQT ansatz} :  The Hamiltonian parameters, $\Vec{h}$, are encoded in the angles of the single qubit parametrized gates through a linear transformation. This along with the linear implementation of CNOTs forms the encoding layer.  The output state from the encoding layer is further processed by the parametrized HVA. The gates in HVA are chosen according to the Hamiltonian whose Gibbs states needs to be prepared.
    An example HVA for a transverse field Ising model is shown, where $XX = e^{-i\theta\sigma_{i}^{x}\sigma_{i+1}^{x}}$. $l$ represents the number of encoding or processing layers. After the final processing layer, measurement on the system qubits yield a density matrix, $\tilde{\rho} (\vec{\omega},\vec{\phi},\vec{\theta})$, that is further optimized to approximate the true Gibbs state of the system. The number of ancillas are taken to be same as the number of system qubits.}
    \label{meta_vqt_ansatz}
\end{figure*}
\section{Proposed Methodology}
In this section, we introduce the finite-temperature collective optimization using two meta-algorithms developed for Gibbs state preparation: the Meta-VQT and the NN Meta-VQT.
\subsection{Finite-temperature collective optimization } \label{meta and nn meta}
In classical machine learning, collective optimization or meta-learning (learning to learn) involves training an outer (or meta) algorithm that updates the inner learning algorithm. This inner algorithm, used for tasks like image classification defined by a dataset and objective, is refined by the meta-algorithm to enhance the model’s performance on broader objectives, such as improving generalization or accelerating the learning speed of the inner algorithm.
Meta-learning also draws on prior experience to handle new tasks more effectively \cite{classical_ml_meta_learning_1,classical_ml_meta_learning_2,classical_ml_meta_learning_3,meta_learning_workflow}.
A schematic diagram that summarizes the meta-learning algorithm is shown in Fig.~\ref{schematics meta}.

Recently, meta-learning has attracted significant attention from the quantum machine learning community as well.
It finds applications in tasks such as generalization of ground states of parametrized Hamiltonian (Meta-VQE and Neural Network (NN) Meta-VQE or quantum circuit learning) \cite{meta-vqe,nn-meta-vqe,coll_opt_vqe1,coll_opt_vqe2}, training algorithms that does not rely on gradient computation \cite{quantum_frj_meta_learning} and training classical neural networks to assist in the quantum learning process \cite{qml_meta_learning_1,qml_meta_learning_2}.

In this work, we introduce the finite-temperature versions of Meta-VQE and NN Meta-VQE, known as Meta-VQT and NN Meta-VQT, which generates the Helmholtz free energy profile, $G(\Vec{h})$, and prepares the corresponding quantum Gibbs state, $\rho(\Vec{h})$, of a parametrized Hamiltonian, $H(\Vec{h})$.
The workflow of our proposed Meta-VQT/NN Meta-VQT is shown in Fig.~\ref{vqt_meta_workflow}(b), where the blue boxes denote modules implemented on a classical computer, while the orange boxes indicate modules that become classically intractable as system size increases and are therefore intended for implementation on a quantum computer.
The finite-temperature meta-algorithms are developed as  generalized versions of the VQT algorithm explained in Sec.~\ref{VQT section}.
Here, the PQC, $\ket{\psi(\Vec{\theta},\Vec{\phi},\Vec{\omega})}$ (or neural network) is trained for multiple discrete input parameters ${\Vec{h}_{train}} \in [\Vec{h}_{1},\Vec{h}_{2},...,\Vec{h}_{n}]$ of the Hamiltonian, $H(\Vec{h})$, so that for the optimal ansatz parameters, $(\Vec{\theta}^*,\Vec{\phi}^*,\Vec{\omega}^*)$ (or $\vec{\phi}_{nn}^*)$, the quantum circuit is able to generate quantum Gibbs state for Hamiltonian parameters $\Vec{h}_{test} \notin {\Vec{h}_{train}}$.
Suppose that $\vec{h}$ is a single-valued vector $h$ and $h_{train} \in (h_{1},h_{2},.....h_{n})$ is the training set with $h$ values arranged in ascending order. Then the meta-algorithms perform well for $h_{test} \notin h_{train}$ and $h_{1} \leq h_{test} \leq h_{n}$, whereas outside this range, its performance tends to decline.
The number of distinctive trainable angles depends on the type of meta-algorithm, size of $\vec{h}$, the type of encoding and the class of PQC used for training.

Since, the target Gibbs state is usually mixed, additional ancilla qubits are required to prepare quantum Gibbs state that we trace out before measurement.
Having equal number of ancilla and system qubits provides the best performance.
However, for simpler Hamiltonians, fewer ancilla qubits also demonstrates high accuracy.

The loss function used is the sum of Helmholtz free energy corresponding to each parameter in $\Vec{h}_{train}$.
For system sizes considered here, where the number of qubits ranges from $n=2 ~\text{to}~8$, the Helmholtz free energy is evaluated on a classical computer after measuring the Gibbs state, $\rho(\vec{h}) \approx \tilde{\rho} (\Vec{\theta}) = \text{Tr}_{\text{ancilla}} \left(\ket{\psi(\Vec{\theta}^*,\Vec{\phi}^*,\Vec{\omega}^*)}\bra{\psi(\Vec{\theta}^*,\vec{\phi}^*,\vec{\omega}^*)}\right)$, from the quantum circuit.
For larger system sizes, where the size of the Gibbs state becomes too large to be stored in a classical computer, the Helmholtz free energy itself is evaluated from the quantum circuits using any of the methods mentioned in Appendix~\ref{Gibbs free energy evaluation}.
All our methods for meta-algorithms remain independent of how we evaluate the Helmholtz free energy.

The following sections outline the details regarding the Meta-VQT and NN Meta-VQT algorithms.
\subsubsection{\textbf{Meta-Variational Quantum Thermalizer (Meta-VQT)}}\label{meta section}
For the parametrized spin Hamiltonian, $H(\Vec{h})$, where $\Vec{h} = (h^{1}, h^{2}, \ldots h^{p})$ represents a vector comprising $p$ parameters of the Hamiltonian that define the interaction strength and/or field strength, consider a discrete finite set of parameters defined by $\Vec{h}_{\text{train}} \in [\Vec{h}_{1}, \Vec{h}_{2}, \ldots, \Vec{h}_{m}]$.
The first step in the Meta-VQT involves encoding the parameters of the Hamiltonian in $\Vec{h}_{train}$ in the quantum circuit, see pink colored box Fig. \ref{vqt_meta_workflow}(b).
This encoding layer is chosen to be a SU2 hardware efficient ansatz. This ansatz is constructed using native single-qubit and two-qubit gates, typically following the SU(2) group operations, and is optimized to align with the physical connectivity of the quantum processor \cite{HEA}.
The number of ancillas in the encoding layer is considered same as that of the system qubits.
The encoding is done using a linear transformation in the angles of the RZ and RY quantum gates, as shown in Fig. \ref{meta_vqt_ansatz}, followed by a linear implementation of CNOTs.
The variables $\Vec{\omega}$ and $\Vec{\phi}$ are the trainable parameters in the encoding layer. If $l$ is the number of encoding layers, then the total number of trainable parameters in the encoding layer is $4 \times (\text{n} + \text{n}_{\text{ancilla}}) \times l \times \text{size of}~\Vec{h}$.

It is proven in several works that the SU2 hardware-efficient ansatz alone in VQT remains insufficient for simulating the quantum Gibbs state of complex Hamiltonians, especially near the quantum critical regime \cite{simulation_quantum_critical}.
Several ansatze such as the product spectrum ansatz \cite{product_spectrum_ansatz} and the Hamiltonian variational ansatz (HVA) \cite{simulation_quantum_critical,verdon_vqt,HVA1,HVA2,HVA3} have been proposed to efficiently simulate the quantum Gibbs state of these complex many-body Hamiltonians.
With this motivation, the quantum state from the encoding layer is further processed using HVA with trainable parameters followed by a linear implementation of CNOTs, as shown in Fig.~\ref{meta_vqt_ansatz}.
The total number of qubits in the HVA layer is also taken as $\text{n} + \text{n}_{\text{ancilla}}$.
The HVA ansatz, $U(\Vec{\theta}) = e^{-i \vec{\theta} H(\Vec{h})}$
have parametrized $1$ and $2$ qubit quantum gates chosen according to the Hamiltonian terms under consideration.
The example of HVA for TFIM is shown in Fig.~\ref{meta_vqt_ansatz}.
The number of trainable parameters, $\vec{\theta}$, in HVA depends on the terms in the Hamiltonian whose Gibbs state needs to be prepared.
\begin{figure*}[!tbh]
    \centering
    \includegraphics[scale=0.2]{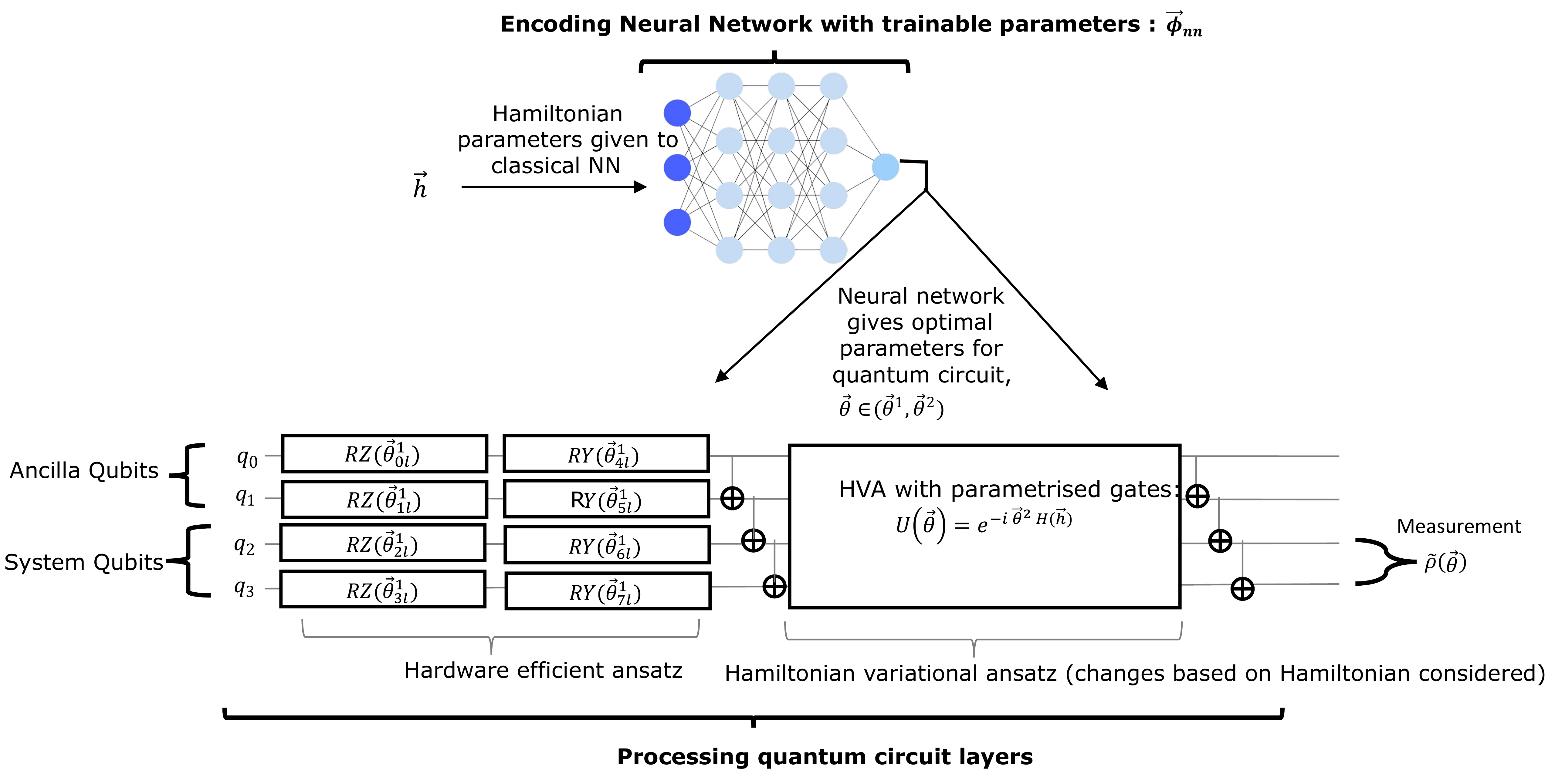}
    \caption{ \textbf{NN Meta-VQT ansatz} : In NN Meta-VQT, the neural network encodes the Hamiltonian parameters, $\vec{h}$, and provides the circuit parameters, $\vec{\theta}$, as output. The quantum circuit is constructed out of both hardware-efficient and HVA as shown above. Similar to Meta-VQT, after this final processing layer, measurement on the system qubits yield a density matrix, $\tilde{\rho} (\vec{\theta})$, that is further optimized to approximate the true Gibbs state of the system.}
    \label{NN Meta VQT ansatz}
\end{figure*}

The density matrix, $\tilde{\rho}(\Vec{\omega},\vec{\phi},\Vec{\theta},\Vec{h})$, measured by tracing out the ancillas is used to evaluate the Helmholtz free energy, $G(\Vec{h})$, of $H(\Vec{h})$.
The trainable parameters, $\Vec{\omega},\Vec{\phi},\Vec{\theta}$, in the encoding and processing layers are initialized randomly at the first iteration.
With the same random initialization, all the parameters in $\Vec{h}_{\text{train}}$ are encoded step by step and corresponding $\tilde{\rho}(\Vec{\omega},\vec{\phi},\Vec{\theta},\Vec{h})$ and $G(\Vec{h})$ are evaluated, see Fig.~\ref{vqt_meta_workflow}(b).
The global loss function, $G_{\text{total}}$, is then evaluated as,
\begin{multline}\label{global_cost}
    G_{\text{total}} = \sum_{\Vec{h}_{i} \in \vec{h}_{\text{train}}} G(\vec{h}_{i}),\\
=  \sum_{\Vec{h}_{i} \in \vec{h}_{\text{train}}} \text{Tr}[H(\vec{h}_{i})\Tilde{\rho}(\Vec{\omega},\vec{\phi},\Vec{\theta},\Vec{h}_{i})] + \\
\frac{1}{\beta}\text{Tr}[\Tilde{\rho}(\Vec{\omega},\vec{\phi},\Vec{\theta},\Vec{h}_{i})\log \Tilde{\rho}(\Vec{\omega},\vec{\phi},\Vec{\theta},\Vec{h}_{i})].
\end{multline}
The $G_{\text{total}}$ is minimized, and the trainable parameters are updated using the classical optimizer ADAM until convergence is reached.
Once converged, the quantum state measured at optimum parameters, $\Vec{\omega}^*,\Vec{\phi}^*,\Vec{\theta}^*$, of the quantum circuit, by tracing out ancillas, is expected to be close to the true quantum Gibbs state of the given Hamiltonian at parameters $\Vec{h}_{j} \in \Vec{h}_{\text{train}}$ as well as $\Vec{h}_{j} \notin \Vec{h}_{\text{train}}$.

The chosen ansatz plays a crucial role in accurately representing thermal correlations in many-body systems. The Gibbs state of many-body quantum systems can display growing mutual information with subsystem size, driven by both the Hamiltonian parameters and temperature, and particularly pronounced near quantum critical points. Through numerical simulations we observe that our composite ansatz, which interleaves hardware-efficient (HEA) and Hamiltonian variational (HVA) layers with CNOT gates, successfully captures this behavior with high fidelity.
This hybrid structure enables faithful approximation of complex Gibbs states across diverse values of $\vec{h}$ and $\beta$.
\subsubsection{\textbf{Neural Network Meta-Variational Quantum Thermalizer (NN Meta-VQT)}}\label{nn meta section}
\begin{figure*}[!tbh]
    \centering
    \includegraphics[width=1\textwidth, height = 175pt]{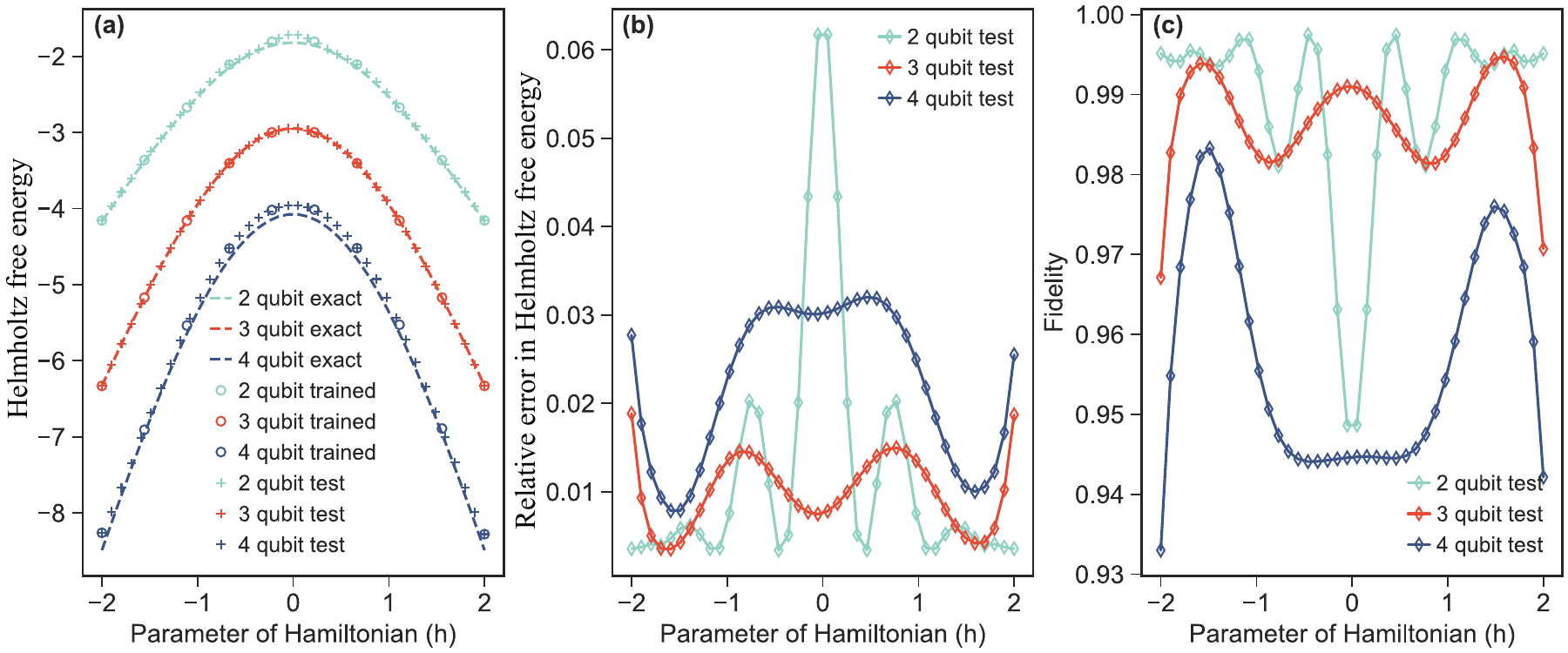}
    \caption{\textbf{Gibbs state preparation of TFIM using Meta-VQT} : The results of Meta-VQT algorithm run for different qubit sizes of TFIM is shown. The value of $\beta$ is chosen to be $1$. (a) shows the exact Helmholtz free energy (dashed) evaluated in the interval $h \in [-2,2]$, and the one calculated after training from the variationally prepared Gibbs state for the parameters which are present (open circles) and absent (cross mark) in the training set. The relative error between exact Helmholtz free energy and the Helmholtz free energy evaluated using variationally prepared Gibbs state at $h_{test}$ with trained parameters of Meta-VQT is shown in (b). \rev{The low relative errors show accurate Gibbs state preparation at held-out test parameters within the domain spanned by the training set.} This is also evident from (c) where the fidelity between exact Gibbs state and Gibbs state measured from the trained Meta-VQT for test parameters gives value between $0.93$ to $1$. The number of encoding and processing layers chosen for system sizes $n=2,3,4$ are $[2,2],[3,3]~\text{and}~[4,4]$ respectively.}
    \label{results_tfim_metavqt}
\end{figure*}

Neural Network Meta-VQT (NN Meta-VQT) builds on the previous variant by adding a classical neural network to the quantum circuit as shown in Step 3 (green colored box) of Fig.~\ref{vqt_meta_workflow}(b).
The neural network receives the Hamiltonian parameters from $\vec{h}_{train}$, as input and provides the parameters of the quantum circuit, $\vec{\theta}$, as output, essentially taking the place of the encoding layer.
This quantum circuit that takes the parameters from neural network forms the processing layer that gives the approximation to Gibbs state of the Hamiltonian.
Unlike Meta-VQT, here the trainable parameters, $\vec{\phi}_{nn}$, are present in the neural network rather than in the quantum circuit.
The number of trainable parameters (size of $\vec{\phi}_{nn}$) depends on the type of neural network, number of hidden layers considered and the type of processing layers included in the ansatz.
Also, the number of classical trainable parameters are greater in number than the trainable parameters in Meta-VQT, thus providing an additional layer of complexity in mapping Hamiltonian parameters to the quantum circuit \cite{nn-meta-vqe}.
At the end of training, the neural network optimizes the mapping between the Hamiltonian parameters $H(\Vec{h})$ and the necessary parameters of the quantum circuit required to generate the Gibbs state.

 The choice of processing layer considered here includes SU2 hardware-efficient ansatz followed by HVA, as shown in Fig.~\ref{NN Meta VQT ansatz}.
The neural network architecture considered here includes $3$ fully-connected hidden layers with sigmoid activation function and a linear layer at the output that matches the dimensions of the parameterized quantum circuit.
  During training, the global loss function, $G_{\text{total}}$ in Eq.~\eqref{global_cost}, which is obtained by summing over Helmholtz free energy corresponding to each of the parameters in the training set, is minimized until convergence.  \\

\rev{\textit{Key Insight: Meta-VQT and NN Meta-VQT provide two complementary approaches to parameter-conditioned Gibbs state preparation. Meta-VQT encodes the Hamiltonian parameters directly within the quantum circuit, while NN Meta-VQT learns a classical mapping from Hamiltonian parameters to quantum-circuit parameters. Both leverage a collective free energy objective across the finite training set, enabling a unified state preparation strategy across multiple Hamiltonian instances.}}
\begin{figure*}[!tbh]
    \centering
    \includegraphics[width=1\textwidth, height = 175pt]{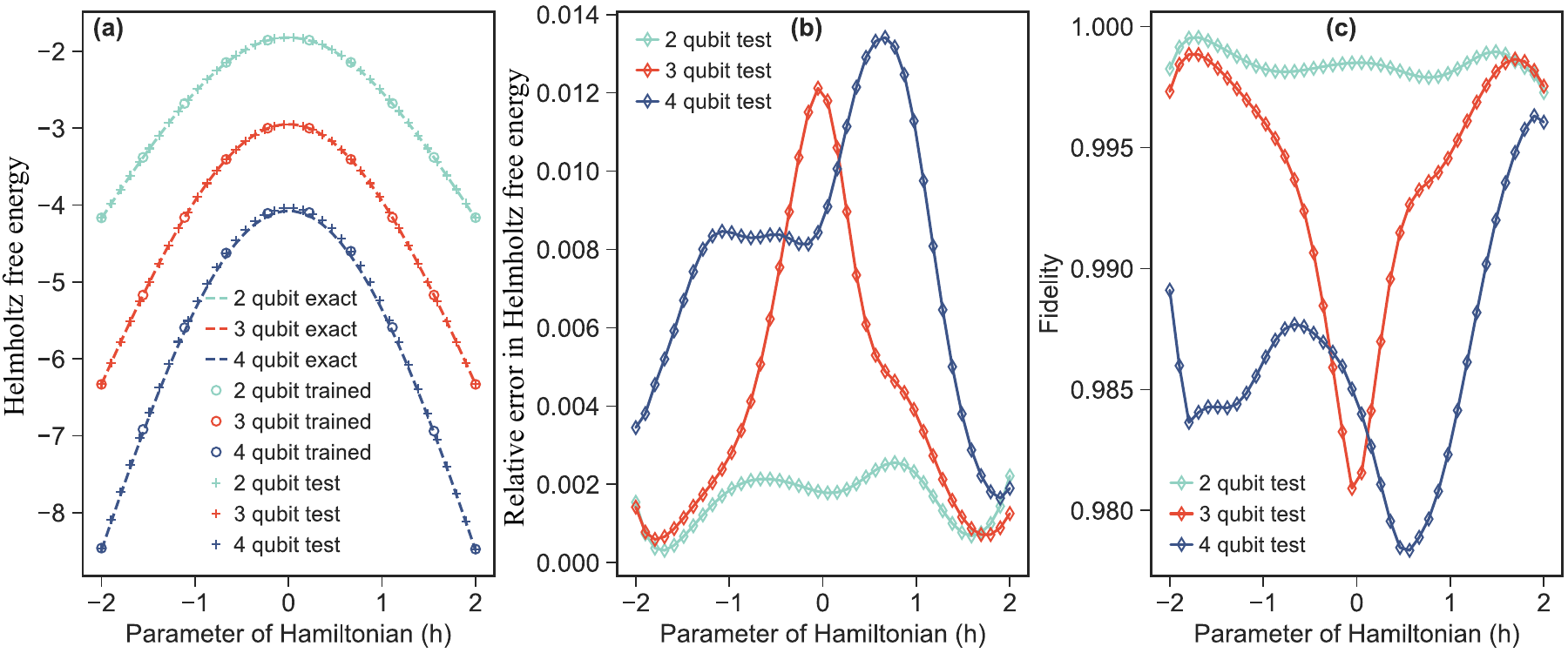}
    \caption{\textbf{Gibbs state preparation of TFIM using NN Meta-VQT} : The results of NN Meta-VQT algorithm for different qubit sizes of TFIM is shown. The value of $\beta$ is chosen to be $1$. (a) shows the exact Helmholtz free energy (dashed) evaluated in the interval $h \in [-2,2]$, and the one calculated after training from the variationally prepared Gibbs state for the parameters which are present (open circles) and absent (cross mark) in the training set. The relative error between exact Helmholtz free energy and the Helmholtz free energy evaluated using variationally prepared Gibbs state at $h_{test}$ with trained parameters of NN Meta-VQT is shown in (b). \rev{The low relative errors show accurate Gibbs state preparation at held-out test parameters within the domain spanned by the training set.} This is also evident from (c) where the fidelity between exact Gibbs state and Gibbs state measured from trained NN Meta-VQT ansatz for test parameters gives value between $0.98$ to $1$ for $n=2,3$ and $4$. The number of SU2 hardware efficient and HVA ansatz layers in the processing part is chosen as $[2,2],~[3,3]$ and $[4,4]$ for system sizes $n=2,3$ and $4$ respectively.}  \label{results_tfim_nnmetavqt}
\end{figure*}
\begin{figure}
    \centering
    \includegraphics[width=1\linewidth, height = 136pt]{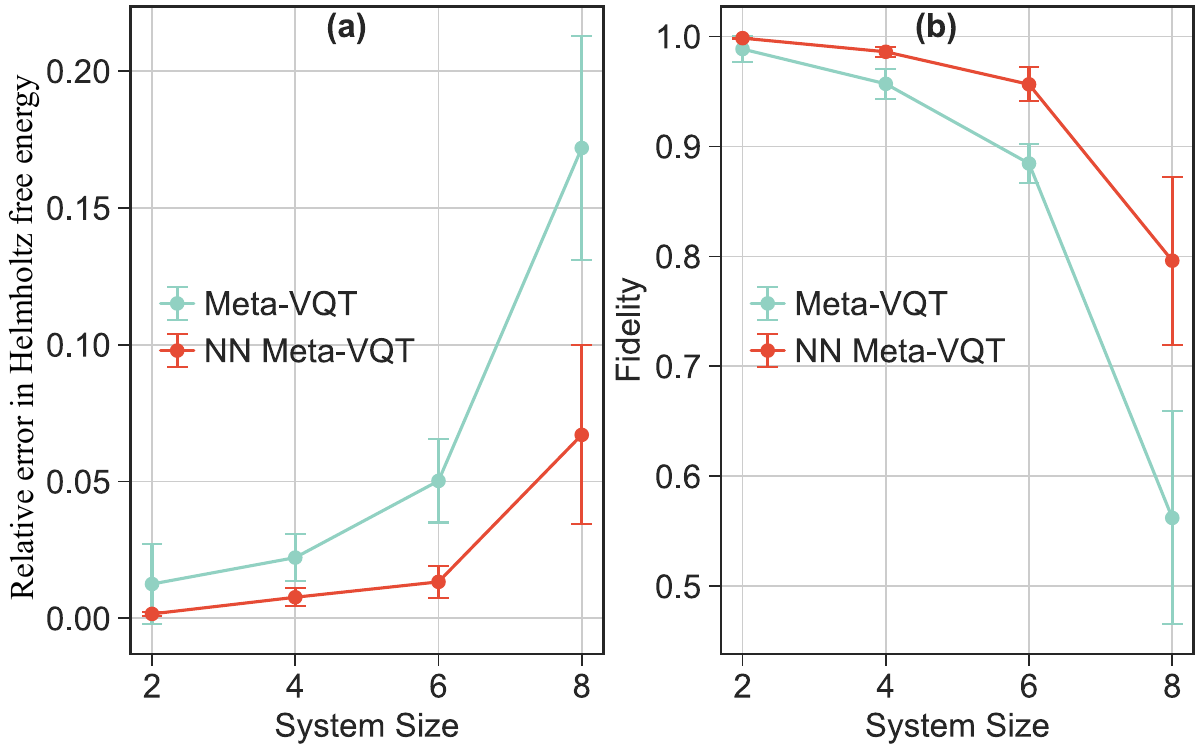}
    \caption{\textbf{Scaling of Meta-VQT and NN Meta-VQT:}(a) shows the mean value with standard deviation of relative error between the Helmholtz free energy evaluated for values of $h$ in $h_{test}$ from the trained meta-algorithms and the exact one for system sizes $2,4,6$ and $8$ qubits. The relative error increases for both Meta-VQT and NN Meta-VQT as system size increases to $8$, with NN Meta-VQT exhibiting lower value than Meta-VQT. (b) shows the mean value and standard deviation of fidelity evaluated between prepared and true Gibbs state at $h_{test}$ as a function of system size. The fidelity decreases for both Meta-VQT and NN Meta-VQT as system size increases to $8$, with a slighltly higher value for NN Meta-VQT than Meta-VQT for larger system sizes. For $n$ qubit Hamiltonian, the number of encoding (SU2 hardware efficient) and processing (HVA) layers considered for Meta-VQT (NN Meta-VQT) is $[n,n]$. The test set, $h_{test}$, over which the mean and standard deviation in relative error in Helmholtz free energy and fidelity is evaluated are $40$ uniform points between $[-2,2]$. }
    \label{scaling results}
\end{figure}
\section{Numerical Results} \label{results meta and nn meta}
This section provides results from the numerical experiments where the Meta-VQT and NN Meta-VQT algorithms are used for preparing Gibbs state of different types of spin Hamiltonians.
The pseudocode for deploying the Meta and NN Meta-VQT is shown under Algorithm \ref{Meta-VQT pseudo} and Algorithm \ref{NN Meta-VQT pseudo} in Appendix \ref{pseudoalgo}.
\subsection{Preparing Gibbs state of TFIM}
In this section, Meta and NN Meta-VQT algorithms are deployed to generate the quantum Gibbs state of a parametrized transverse field ising model (TFIM), $H(h) = -J\sum_{i=1}^{N-1} \sigma_{i}^{x}\sigma_{i+1}^{x} - h\sum_{i=1}^{N}\sigma_{i}^{z}$, where the interaction strength $J=1$ and the Hamiltonian is parametrized as a function of transverse field, $h$.
The inverse temperature $\beta$ is fixed as $1$, the number of ancillas is considered the same as the number of system qubits, and the initial learning rate is taken as $\eta = 0.01$ for Meta-VQT and $0.001$ for NN Meta-VQT.
The first set of experiments are run on three different system sizes $n = 2,3~\text{and}~4$.
The set of Hamiltonian training parameters $h_{train}$ considered are $10$ uniformly spaced points between $-2$ and $2$.
The number of encoding and processing layers is chosen as $[2,2],~[3,3],~\text{and}~[4,4]$ for system sizes $n=2,3~\text{and}~4$, respectively.
For all system sizes in NN Meta-VQT, the ansatz configuration has a feed forward neural network consisting of one input, one hidden and two output layers.
The output layers from these neural networks provide angles for SU2 hardware efficient and  HVA layers.
For both Meta and NN Meta-VQT, the test set of Hamiltonian parameters, $h_{test}$ is considered as $40$ uniformly spaced points between $-2$ and $2$. This held-out set also has points that are not present in the $h_{train}$.

Figure~\ref{results_tfim_metavqt}(a), shows the Helmholtz free energy evaluated from the approximate Gibbs state measured from the trained Meta-VQT ansatz for different system sizes.
The open circle corresponds to the variational Helmholtz free energy evaluated at $h_{train}$ while the cross marker represents the one evaluated for $h_{test}$.
The true Gibbs state evaluated analytically in the interval $[-2,2]$ is indicated as dashed lines.
The relative error between true Helmholtz free energy and the one evaluated from the approximate Gibbs state for the test parameter set is shown in Fig. \ref{results_tfim_metavqt}(b).
From Figs. \ref{results_tfim_metavqt} (a) and (b), we can conclude that the trained Meta-VQT can generate the Helmholtz free energy profile with small relative error for all system sizes for the test set with just $10$ training points.
Since the system size considered is relatively smaller, the quality of the prepared Gibbs state is further evaluated by analyzing its fidelity with the exact one as shown in Fig. \ref{results_tfim_metavqt}(c).
The fidelity stays in the range between $0.93$ to $1$, with a lower fidelity for larger system size.

Figure~\ref{results_tfim_nnmetavqt} shows the same set of numerics run with the NN Meta-VQT ansatz.
\rev{From Fig.~\ref{results_tfim_nnmetavqt}(a)--(c), NN Meta-VQT accurately generates the Helmholtz free energy profile and Gibbs states at test values omitted from, but lying within the interval spanned by, the training set.}
The relative error in Helmholtz free energy(fidelity) for test set is found to be slightly less (more) for states prepared using NN Meta-VQT than Meta-VQT. A brief discussion of the classical optimization overhead associated with meta-training is provided in Appendix~\ref{Classical_optimization_overhead}.

\subsubsection{Finite-size scaling of meta-algorithms}
The performance of these meta-algorithms as we scale the system size from $2$ to $8$ qubits is shown in Fig.~\ref{scaling results}(a) and (b).
Figure~\ref{scaling results}(a) plots the mean value with standard deviation of the relative error between the Helmholtz free energy evaluated from the trained meta- algorithms and the exact one for values of $h$ in $h_{test}$ as system size increases from $2$ to $8$ qubits.
Figure~\ref{scaling results}(b) shows the mean value with standard deviation of fidelity evaluated between the prepared and true Gibbs state as a function of system size for values of $h$ in $h_{test}$.
As the system size increases, the relative error (fidelity) increases (decreases), with a noticeably larger error bar observed for $N=8$. This behavior may be correlated with the systematic decrease in the purity of the exact Gibbs states as the system size grows. In particular, larger systems correspond to increasingly mixed thermal states, which can be more challenging for the chosen ansatz to represent accurately.
Also, NN Meta-VQT performs slightly better than Meta-VQT for the two to eight qubit cases studied.

\rev{This trend indicates that the fixed $[n,n]$ HEA--HVA configuration may not be optimal at every system size. The observed increase in layer count is a property of the ansatz and training protocol used here, not an intrinsic scaling requirement of meta-learning. Hardware-efficient ansatze can be expressive, but their trainability depends on depth, connectivity, initialization, cost locality, noise, and Hamiltonian structure \cite{HEA,barren_plateu1,barren_plateu2,barren_plateu3}. Alternatives worth systematic comparison at fixed target accuracy include adaptive or symmetry-preserving Hamiltonian ansatze, product-spectrum constructions \cite{product_spectrum_ansatz}, and dynamic parameterized circuits \cite{baren_plateau_free_dynamic_pqc}.}

\subsubsection{Meta-algorithms for better initialization} \label{better_initialization}
\begin{figure}
    \centering
    \includegraphics[width=1\linewidth, height = 136pt]{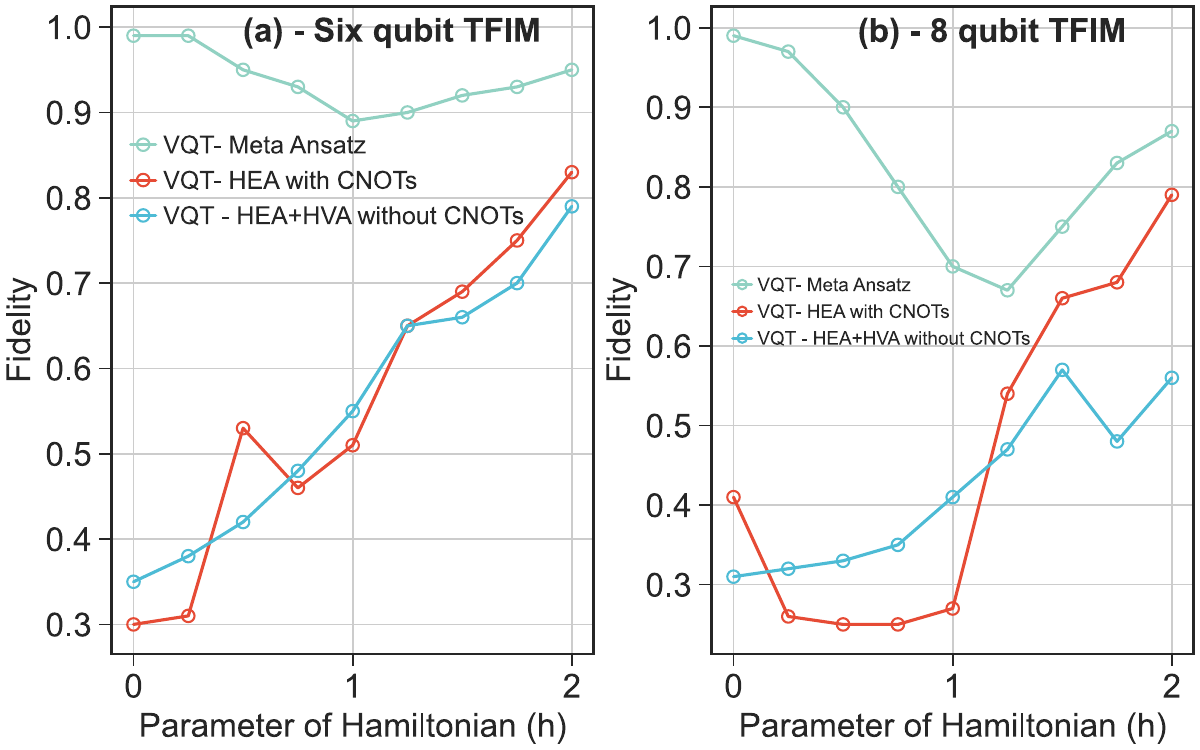}
    \caption{\textbf{VQT initialization with meta-trained ansatz:} Fidelity between the exact and variationally prepared Gibbs states for system sizes (a) $n=6$ and (b) $n=8$ qubits, using different implementations of VQT. The meta-trained HEA+HVA ansatz with CNOTs (cyan color) consistently outperforms other approaches, demonstrating the benefit of meta-learning-based initialization in achieving higher fidelity Gibbs state preparation.}
    \label{better initialization}
\end{figure}

Though the performance tends to decrease at larger qubit sizes ($n \geq 6$), our trained meta-algorithms serve as effective initializations for VQT, particularly when optimizing the Gibbs state with respect to a single Hamiltonian parameter.
To demonstrate this, we employ a VQT approach using our HEA+HVA ansatz interleaved with CNOT gates, initialized with parameters obtained via meta-training, and optimize the Helmholtz free energy corresponding to values of the transverse field parameter $h \in [0, 2]$. We consider system sizes of $n = 6$ and $n = 8$ qubits.
For comparison, we evaluate two alternative approaches:
(i) the HEA ansatz with a linear CNOT layout, a single ancilla and parametrized $R_y$ rotations applied to each of the system and ancilla qubits, as proposed in Wang et al.\cite{wang2005variational}, using Helmholtz free energy as the cost function; and
(ii) the VQT approach used in Verdon et al.\cite{verdon_vqt}, Johannes Selisko et al.\cite{HVA2}, and related works, which combines two PQCs, for instance an HEA followed by an HVA circuit. In the latter method, intermediate measurements from the first circuit are used to compute the Von Neumann entropy, while the second circuit evaluates the Hamiltonian expectation value. Both quantities are used to construct the Helmholtz free energy, and no ancillas are employed. A brief explanation on this work is also provided in Section ~\ref{Gibbs free energy evaluation}.
For the meta-trained VQT and the HEA+HVA ansatz without CNOTs in (ii), we use $[n,n]$ layers. For the HEA+CNOT method in (i), we use $n$ layers of HEA.
Figure~\ref{better initialization}(a) and (b) show the fidelity between the exact and variationally prepared Gibbs states for $n = 6$ and $n = 8$ qubits using three different methods.
\rev{For the six and eight qubit instances tested, the meta-trained initialization leads to significantly higher fidelity than the two randomly initialized VQT baselines considered here. This establishes a finite-size warm-start advantage for the example considered here.}
For the two alternative approaches in (i) and (ii), parameters are initialized randomly. While the meta-learning framework incurs additional computational cost due to the meta-training phase and associated optimization steps, this trade-off results in substantially improved Gibbs state preparation accuracy compared to the alternative methods with random initialization.

Additional numerical results demonstrating that the Meta-VQT and NN Meta-VQT models capture the physical features of the Gibbs state are presented by comparing the purity of the variationally prepared Gibbs state with that of the exact Gibbs state, as shown in Fig.~\ref{purity_meta_nnmeta}(a), (b), and (c) in Appendix~\ref{purityplots}, while Appendix~\ref{results_outsidetraining} further reports the accuracy of states prepared using trained meta-algorithms for points that lies outside the interval of training set. A comparative performance analysis between the proposed meta-algorithms and the existing VarQITE-based Gibbs state preparation is presented in Appendix~\ref{heisenberg} using a $2$ qubit Heisenberg model with all field terms included, where the meta-algorithms consistently outperform VarQITE even for this small-scale Hamiltonian. Furthermore, Appendix~\ref{complexity} provides numerical evidence that the precision of the meta-algorithms remains robust regardless of the number of non-commuting terms in the Hamiltonian, in contrast to VarQITE, whose precision deteriorates as the number of non-commuting terms increases.

\subsubsection{Meta-VQT versus NN Meta-VQT}
From the results obtained for the TFIM (Fig.~\ref{results_tfim_nnmetavqt} \& Fig.~\ref{scaling results}), it is evident that, for the same choice and number of HEA and HVA layers, NN Meta-VQT consistently outperforms Meta-VQT. One possible reason is that NN Meta-VQT contains a larger number of classical trainable parameters compared to Meta-VQT, thereby providing additional expressive power in learning the mapping from Hamiltonian parameters to the corresponding quantum circuit parameters. This enhanced representational capacity is also discussed in the context of variational quantum eigensolvers and ground-state estimation in Ref.~\cite{nn-meta-vqe}.

\rev{Furthermore, Appendix~\ref{results_outsidetraining} examines the behavior of the proposed methods at test points beyond the training interval. NN Meta-VQT maintains higher fidelity than Meta-VQT for the finite cases studied, demonstrating effective local extrapolation beyond the sampled training domain. The fidelity decreases gradually as the test point moves farther from the training interval, particularly for the larger system size, providing a quantitative characterization of the local extrapolation range. Extending this analysis to broader parameter ranges, larger system sizes, and different Hamiltonian families will help further assess the generalization capabilities of the proposed approaches.} In Appendix~\ref{heisenberg}, Table~\ref{table_meta_nn_varqite} compares the accuracies of three different approaches: VarQITE (existing method), Meta-VQT, and NN Meta-VQT, measured in terms of the trace distance between the variationally prepared Gibbs state and the exact Gibbs state. The results show that NN Meta-VQT does not necessarily require a processing HVA layer; even without it, the method is capable of preparing Gibbs states close to the exact ones after meta-training. This suggests that, in certain cases, NN Meta-VQT can achieve comparable or superior accuracy with reduced quantum resource requirements relative to Meta-VQT.

Nevertheless, Meta-VQT remains particularly relevant for Hamiltonians exhibiting finite-temperature phase transitions, such as the Kitaev ring model considered in the following section. Although both algorithms efficiently capture the Gibbs state after meta-training, in several instances NN Meta-VQT offers a slightly more efficient and potentially lower resource route to Gibbs state preparation.\\
\rev{\textit{Key Results : (i) Both Meta-VQT and NN Meta-VQT accurately prepare Gibbs states at
held-out parameter values within the training domain, with NN Meta-VQT
consistently achieving slightly higher fidelity and lower Helmholtz
free energy error.
(ii) Meta-trained parameters provide an effective warm-start for
subsequent VQT optimization, yielding substantially higher fidelity
than randomly initialized VQT for the $6$ and $8$ qubit cases studied.
(iii) NN Meta-VQT can, in some cases, achieve comparable or higher
accuracy with reduced quantum resource requirements.}}

\begin{figure*}[!tbh]
    \centering
    \includegraphics[width=1\linewidth]{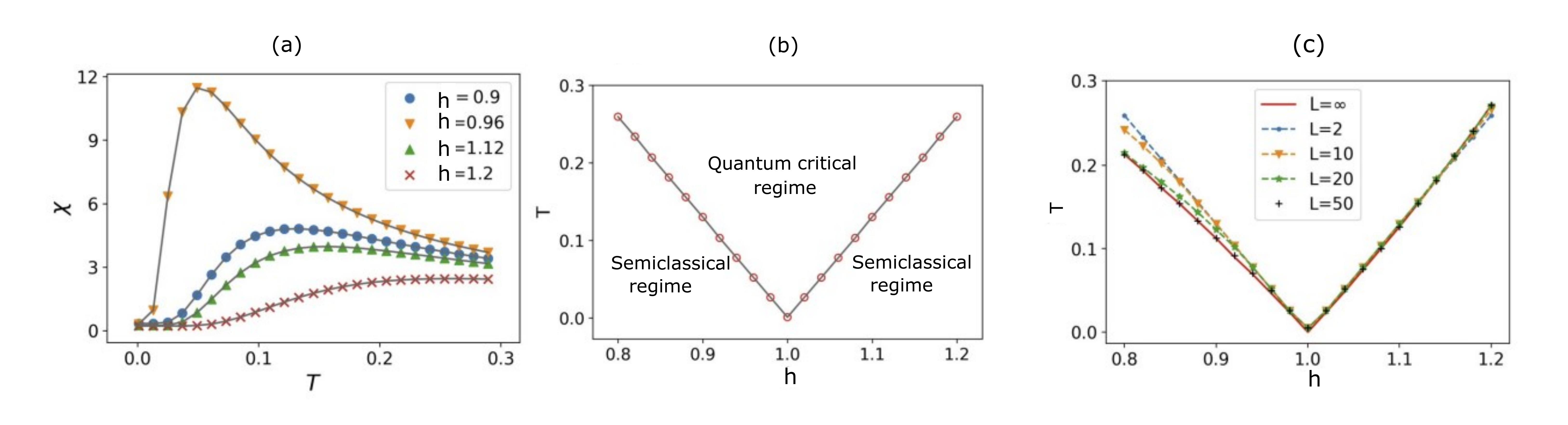}
    \caption{\textbf{Finite-temperature phase diagram of Kitaev ring model} : (a) is the magnetic susceptibility, $\chi$, plotted as a function of temperature, $T$, for different h values. For a particular value of $h$, the temperature at which magnetic susceptibility takes maximum value gives the crossover temperature and thus the phase diagram, which is shown in (b). The calculations in both (a) and (b) correspond to system size in the thermodynamic limit. (c) plots the crossover temperature evaluated for finite system sizes $L=2, 10,20,50$ along with the one in thermodynamic limit, $L = \infty$. From (c), it is evident that smaller system sizes also gives the phase diagram close to that of the one at $L = \infty$. Figure (a) and (b) are adapted from Shi et al.\cite{simulation_quantum_critical} and (c) is adapted from Zhang et al.\cite{kiatev_ring2} .}
    \label{phase_diagram}
\end{figure*}
\begin{figure}
    \centering
    \includegraphics[width=1\columnwidth, height = 136pt]{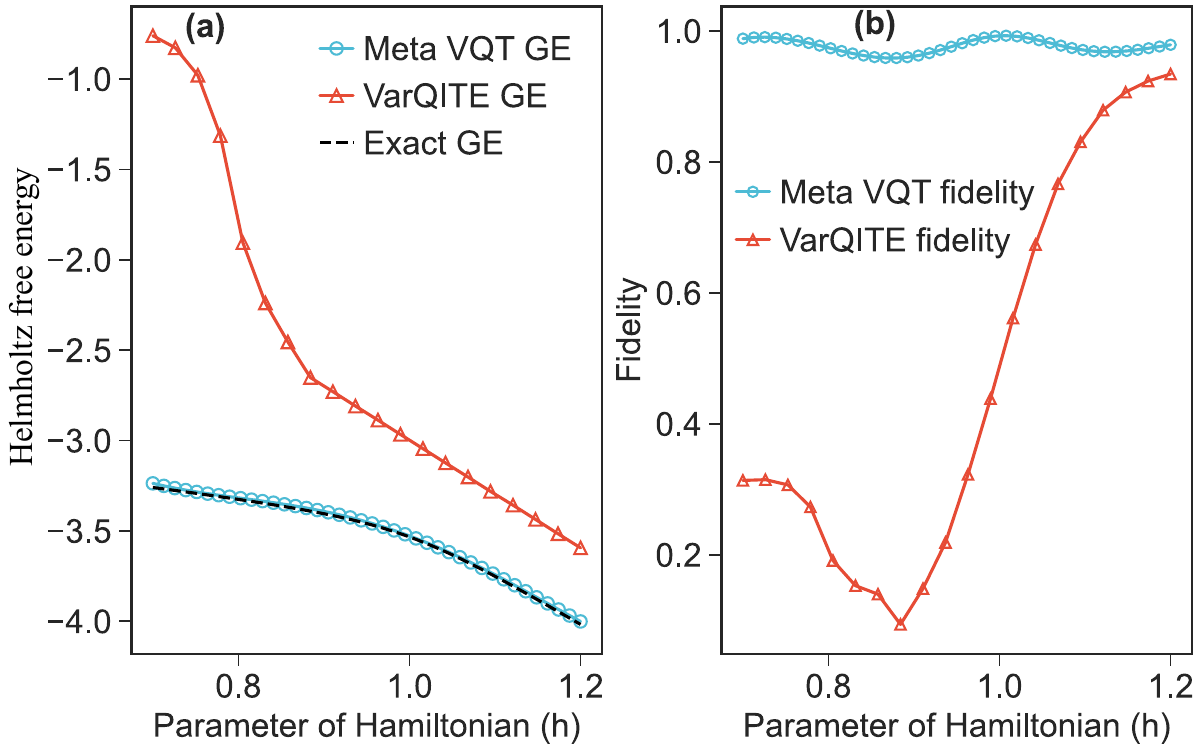}
    \caption{\textbf{Preparation of Gibbs state in different regimes of Kitaev ring model} : (a) shows the plots of Helmholtz free energy evaluated from the approximate Gibbs state prepared using the trained Meta-VQT and VarQITE for different $h$ values of Kitaev ring Hamiltonian. For Meta-VQT the $h$ values considered are $40$ uniform points between $[0.7,1.2]$ including $20$ training points, while for VarQITE it is $20$ uniform points between $[0.7,1.2]$. The black dashed lines correspond to exact Helmholtz free energy. \rev{The trained Meta-VQT generates a more accurate Helmholtz free energy profile than VarQITE at test values omitted from, but contained within, the interval $[0.7,1.2]$.} Additionally, (b) shows the fidelity calculated between the exact Gibbs state and the one approximated using both Meta-VQT and VarQITE. The fidelity for Meta-VQT remains close to one while that of VarQITE depends on the parameter values as well as the regimes of the Hamiltonian. The system size considered is $N=3$. The number of encoding and processing layers for Meta-VQT is $4$ SU2 hardware efficient and $1$ HVA respectively. The depth of SU2 hardware efficient layers in VarQITE is considered to be $5$.}
    \label{kitaev_mo}
\end{figure}
\subsection{Preparing Gibbs state across different regimes of Kitaev ring Hamiltonian}
To establish our second claim which states that Meta-VQT and NN Meta-VQT can efficiently simulate the quantum Gibbs state independent of different regimes of the many-body Hamiltonian, we consider a Kitaev ring (a spinless p-wave superconductor) Hamiltonian,
\eq{ \label{Kitaev ring}
H(h) = -J \sum_{i=1}^{L-1} \sigma_{i}^{x}\sigma_{i+1}^{x} - J \sigma_{1}^{y} P \sigma_{N}^{y} - h \sum_{i=1}^{L} \sigma_{i}^{z},
}
where, $J=1$, $P = \Pi_{i=2}^{L-1} \sigma_{i}^{z}$ is a string operator and $L$ is the system size.
At zero-temperature, the model exhibits a topological quantum phase transition at $h/J = 1$ \cite{pwave_super}.

For finite-temperatures, the Hamiltonian exhibits a temperature crossover behavior as shown in Fig.~\ref{phase_diagram}(b). This is characterized by a smooth and gradual transition between semi-classical and quantum critical regimes as a function of temperature and the Hamiltonian's parameter values \cite{kiatev_ring2}.
The finite-temperature phase diagram in Fig.~\ref{phase_diagram}(b) that shows this crossover behavior is obtained by evaluating the temperature at which magnetic susceptibility, $\chi$, takes maximum value, see Fig.~\ref{phase_diagram}(a).
Zhang et al.\cite{kiatev_ring2} showed that the model exhibits finite-temperature crossover close in range with the infinite system size phase diagram for a wide range of finite system sizes starting from $L=2$, see Fig.~\ref{phase_diagram}(c).
\begin{figure*}[!tbh]
    \includegraphics[scale=0.62]{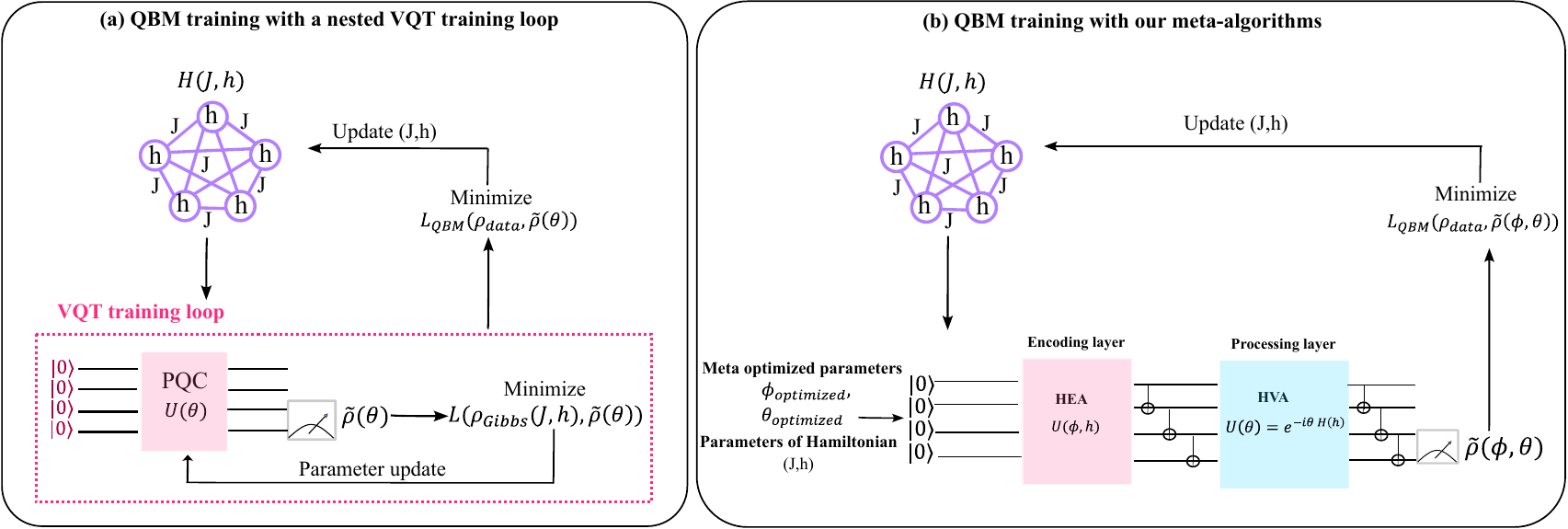}
    \caption{\textbf{Overview of conventional and meta-algorithm-based QBM training : }
    (a) Conventional QBM training: To model a target data distribution $\rho_{data}$, the Gibbs state of a model Hamiltonian $H(J,h)$ must be prepared within each iteration of the QBM training loop. This requires a nested optimization loop where circuit parameters $\theta$ are updated separately for each hamiltonian parameter set $(J,h)$. Once the Gibbs state $\tilde{\rho}(\theta)$ is prepared, the QBM loss $L_{QBM}(\rho_{data},\tilde{\rho}(\theta))$ is minimized by updating $(J,h)$. The dotted pink box highlights the expensive inner-loop Gibbs training step where $\theta$ is optimized. (b) QBM training with meta-algorithm: Here, the variational ansatz is designed to encode the Hamiltonian parameters $(J,h)$ directly to the circuit. The ansatz is initialized with pretrained $\phi_{optimized}$ and $\theta_{optimized}$ obtained from the meat-training stage (as described in Fig.~\ref{meta_algorithm_overview}(b)). As a result, Gibbs state preparation no longer requires a separate training loop for each $(J,h)$ and the optimization is performed only over $(J,h)$. This eliminates the need for the inner-loop $\theta$ training shown in the dotted pink box in (a), significantly reducing the QPU calls.}
    \label{QBM_training_overview}
\end{figure*}

Here, we consider a system size of $L=3$ at temperature of $T= 1/\beta = 0.1$.
The Meta-VQT with $6$ total qubits, $4$ encoding and $1$ processing layer is trained for the subsequent generation of quantum Gibbs state as a function of the strength of transverse field $h$.
A total of $20$~$h$ values are chosen uniformly from the interval $[0.7, 1.2]$ for training purposes.
The reason for this particular choice of interval stems from the fact that as $h$ varies from $0.7$ to $1.2$ the model exhibits transition from semi-classical to quantum critical to semi-classical regime for a temperature value of $T=0.1$, see Fig.~\ref{phase_diagram}(b) and (c).
For testing the Meta-VQT algorithm, $40$ uniform $h$ values are taken from the interval between $[0.7,1.2]$.

The blue open circles in Fig.~\ref{kitaev_mo}(a) shows the Helmholtz free energy evaluated from the Gibbs state prepared using trained Meta-VQT as a function of $h~\in~[0.7,1.2]$.
A comparison of this has been made with the Helmholtz free energy evaluated exactly (black dashed lines) and that evaluated from the Gibbs state prepared using VarQITE (red triangle) that has $5$ layers of SU2 hardware efficient ansatz.
The Helmholtz free energy profile generated from the trained Meta-VQT is more accurate to the exact one than the one evaluated from the states prepared using VarQITE.
Moreover, from Fig.~\ref{kitaev_mo}(b), that plots fidelity between the exact Gibbs states and the ones prepared using trained Meta-VQT (blue open circles) and VarQITE (red triangles) we observe that the Meta-VQT is able to prepare Gibbs state with good fidelity in all finite-temperature regimes of the model while that of VarQITE is dependent on the parameter values as well as the finite-temperature regimes of the model.\\
\rev{\textit{Key Results : Meta-VQT accurately prepares Gibbs states across the semiclassical and quantum critical regimes of the finite-temperature Kitaev ring, maintaining good fidelity across the full parameter range. It also reproduces the Helmholtz free energy profile more accurately than the VarQITE benchmark, demonstrating robust parameter-conditioned Gibbs state preparation across different finite-temperature regimes.}}
\vspace{-0.3cm}
\subsection{Application of NN Meta-VQT: Quantum Boltzman machine} \label{nn meta vqt qbm workflow}
In this section, we show one application of our finite-temperature collective optimization algorithm by implementing the NN Meta-VQT in variational QBM workflow.

A QBM relies on preparing finite-temperature quantum Gibbs states of chosen Hamiltonians as a core subroutine. Samples measured from these states are then used to model the target data.
An overview on classical and quantum Boltzmann machines are provided in Appendix \ref{overview_qbm}.
In the traditional variational QBM training workflow (see Fig.~\ref{QBM_training_overview}(a)), once the Hamiltonian and its parameters are initialized, the Gibbs state is prepared using variational methods such as the VQT \cite{verdon_vqt} which employ hybrid quantum-classical optimization or VarQITE that deploys imaginary time evolution using a hybrid quantum-classical algorithm\cite{mcardle2019variational, Zoufal_circuit_training}.  Samples from the prepared Gibbs state are used to evaluate the QBM loss function, typically based on Kullback-Leibler (KL) divergence \cite{amin_qbm} or quantum relative entropy \cite{bound_based_qbm2, state_based_1}.

For VQT-based QBM training, if $n_{\text{qbm}}$ are the total epochs required for QBM training and $n_{\text{gibbs}}$ is the average number of optimization steps required for thermal state preparation, then the minimum number of QPU calls required for this algorithm is $n_{\text{qbm}} \times n_{\text{gibbs}}$, assuming that the cost function is evaluated on classical computer and ignoring the shot noise.
This implies an additional nested loop for Gibbs state preparation within the QBM training as shown inside pink dotted box in Fig.~\ref{QBM_training_overview}(a).
Since our protocol enables collective optimization over a parameterized Hamiltonian, it allows the QBM to bypass this nested loop as illustrated in Fig.~\ref{QBM_training_overview}(b).  Consequently, the total number of QPU calls reduces to approximately $n_{\text{qbm}} + n_{\text{gibbs}}$ from $n_{\text{qbm}} \times n_{\text{gibbs}}$.
For VarQITE-based QBM training as well, Gibbs state preparation constitutes a nested loop within the QBM training procedure, since for each update of the QBM Hamiltonian parameters, the imaginary time evolution must be re-executed until convergence. A brief review on VarQITE based Gibbs state preparation is provided in Appendix~\ref{varqite}.

To demonstrate the benefit of using a collective optimization based module in QBM training, we consider generative learning mode of the QBM where the classical probability distribution of a target dataset is modeled by the probability distribution of the Gibbs state of the Hamiltonian.
NN Meta-VQT is used for the Gibbs state preparation and compared with a QBM training that uses VarQITE.
The pseudo algorithm for implementing the NN Meta-VQT based QBM training is shown in Algorithm \ref{variational QBM pseudo}.

The target probability distribution to be learned by the variational QBM is chosen to be $p_{target} = [0.62,0.17,0.17,0.04]$, which corresponds to a random $2$ qubit quantum state.
The model Hamiltonian is considered as the $2$ qubit double parameterized Heisenberg Hamiltonian with all field terms in Eq.~\ref{complex_hamiltonian}.
Since NN Meta-VQT with $4$ processing layers performs the best for this Hamiltonian as observed from Table \ref{table_meta_nn_varqite}, we use the trained ansatz of NN Meta-VQT of this Hamiltonian for the preparation of the Gibbs state.

\begin{figure}
    \centering
    \includegraphics[width=1\columnwidth]{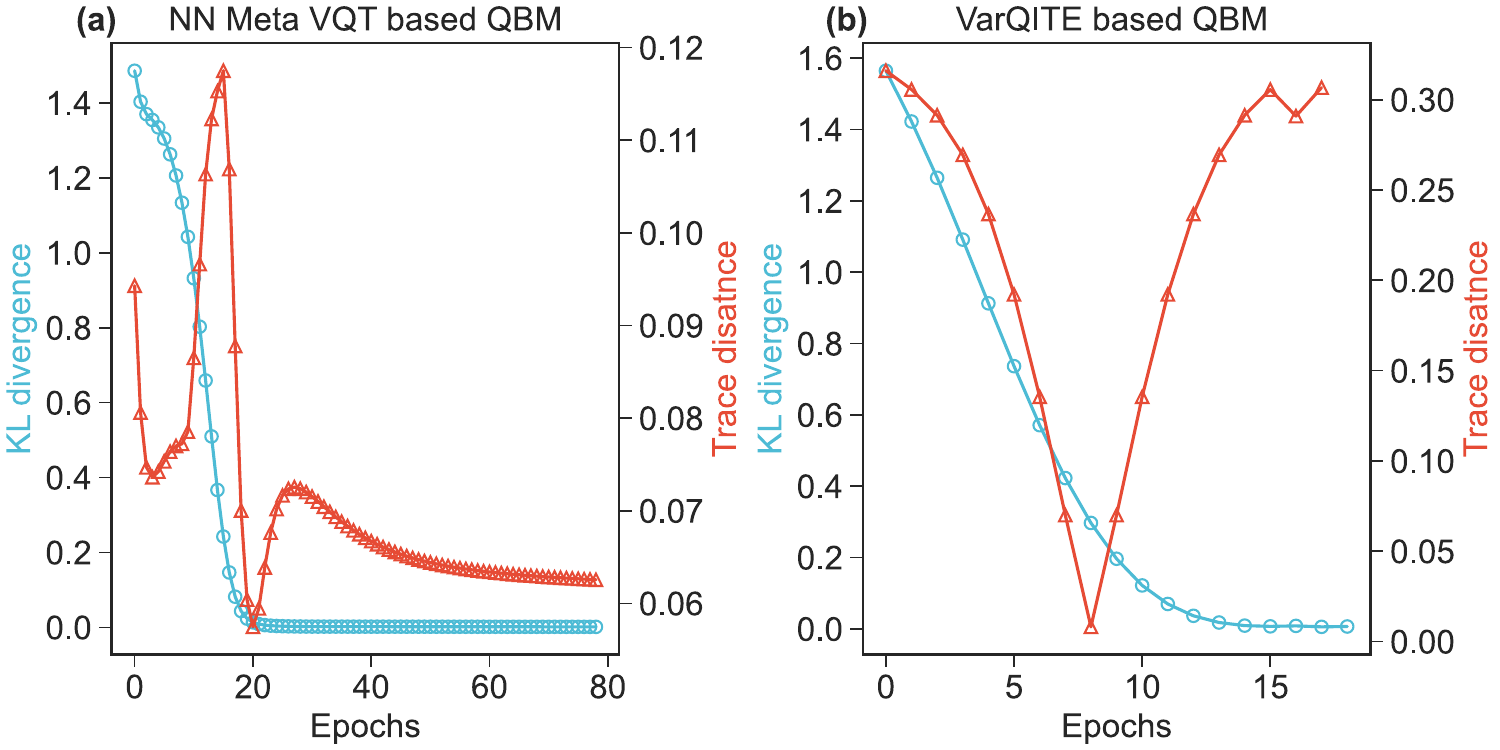}
    \caption{\textbf{NN Meta-VQT in QBM workflow} : (a) and (b) shows the convergence of loss function and the trace distance between exact and prepared Gibbs state using our method of the NN Meta-VQT and existing method of the VarQITE respectively. The KL divergence in (a) converges to $0.0012$ while that in (b) converges to $0.0067$. Additionally, the prepared Gibbs state is found to be more accurate and precise for NN Meta-VQT based QBM than VarQITE based QBM. The number of layers considered for NN Meta-VQT is $4$ SU2 while VarQITE requires $32$ layers of SU2 circuit to prepare Gibbs state with an average value of trace distance $0.22 \pm 0.094$.}
    \label{QBM_results}
\end{figure}

Figure~\ref{QBM_results}(a) and (b) shows that during training the loss function of QBM (blue open circles) converges to a lower value, $0.0012$ when NN Meta-VQT is used for preparing Gibbs state as compared to VarQITE based QBM which converges to a higher value of $0.0067$.
During the training phase, the trace distance between the exact and prepared Gibbs state from NN Meta-VQT and VarQITE (red triangles) is plotted in Fig.~\ref{QBM_results}(a) and (b).
For NN Meta-VQT based QBM, the average trace distance during the training phase is $0.071 \pm 0.013$, and  remains much lower than the one for VarQITE based QBM that has a value of $0.22 \pm 0.094$.
The circuit depth for VarQITE is $32$, which is also much larger than the depth required by NN Meta-VQT to achieve this precision.
The runtime for training NN Meta-VQT is $\approx 40~\text{minutes}$ and with this trained ansatz the QBM training takes $16~\text{seconds}$.
\textit{However, for VarQITE based QBM the total runtime is around $1200~\text{minutes}$ which is $30$ times more than the total runtime of NN Meta-VQT based QBM which takes $\approx 40~\text{minutes}~16~\text{seconds}$}.

\rev{\textit{Key Results : NN Meta-VQT removes the nested Gibbs state optimization loop in QBM training by learning a reusable parameter-conditioned Gibbs state preparation circuit, reducing QPU calls. In the demonstrated 2 qubit benchmark, it also achieves lower QBM loss, higher Gibbs state accuracy, shallower circuits, and substantially shorter runtime than VarQITE, while larger system scaling remains to be established.} }

\section{Conclusions}
\rev{We introduced Meta-VQT and NN Meta-VQT as a transferable, finite-temperature extension of collective variational optimization. The framework learns from Hamiltonian parameters and a collective free energy objective rather than from labeled Gibbs states, and it prepares mixed thermal states rather than only predicting scalar thermal observables. The study combines state preparation, systematic variational benchmarking, warm-start initialization, finite-temperature crossover tests, and an end-to-end QBM application.}

\rev{Across the parametrized TFIM, double parameterized Heisenberg, and Kitaev ring Hamiltonians, we find that Meta-VQT and NN Meta-VQT can learn Gibbs state representations from a finite set of Hamiltonian parameters and generalize to unseen parameters within the training domain. In the TFIM, training with $10$ parameter points provides comparable accuracy across the tested parameter intervals, while the Kitaev ring results show that denser training can be beneficial near finite-temperature crossover regions. For the $2$ qubit double parameterized Heisenberg Hamiltonian, Meta-VQT and NN Meta-VQT prepare Gibbs states with higher precision than VarQITE, while the Kitaev ring results demonstrate that a common meta-trained ansatz can capture thermal state behavior across different finite-temperature regimes without requiring separate ansatz constructions for each regime.}

\rev{We further demonstrated the use of NN Meta-VQT as a Gibbs state preparation module within a variational QBM workflow. By reusing the parameter-conditioned ansatz throughout QBM training, the repeated inner loop Gibbs state optimization required by conventional VQT or VarQITE-based approaches can be avoided. In the two qubit benchmark, the NN Meta-VQT based QBM achieved a better convergence than the VarQITE-based QBM and the average trace distance to the exact Gibbs state for NN Meta-VQT based QBM remained lower during the entire training as compared with VarQITE-based QBM. The total runtime was approximately $40$ minutes and $16$ seconds compared with approximately $1200$ minutes for the VarQITE-based workflow, corresponding to an approximately $30\times$ reduction for this benchmark. These results illustrate the potential of transferable Gibbs state preparation to reduce repeated optimization overhead in downstream thermal quantum learning workflows.}

\rev{Taken together, these results establish a finite-size demonstration of transferable variational Gibbs state preparation and its integration with a downstream QBM learning workflow. At the same time, they raise few questions concerning scalability, generalization, trainability, resource requirements, and hardware implementation. We discuss these open questions and the directions they suggest for future work in the following section.}

\section{Open questions}
There are some challenges that must be addressed while implementing the proposed Meta and NN Meta-VQT algorithms in this work.
The major one arises when the size of the Hamiltonian whose Gibbs state needs to be prepared increases beyond $N \geq 10$.
In this case, the evaluation of exact Helmholtz free energy becomes challenging due to the exponential increase in size of the density matrix that represents the Gibbs state and its memory requirement.
Alternate methods can be used to evaluate the Helmholtz free energy as the ones explained in Appendix ~\ref{Gibbs free energy evaluation}.

\rev{Furthermore, the increasing number of encoding and processing layers and trainable parameters raises questions regarding trainability and hardware resource requirements. A practical challenge in implementing these methods on quantum hardware is the estimation of gradients: parameter-shift methods require additional circuit executions for each trainable parameter, while finite-shot sampling, readout and gate noise, circuit depth, and the number of Hamiltonian instances in the meta-training set can increase the cost and variance of the estimated gradients~\cite{Schuld2019}. Apart from these hardware related costs there are challenges from barren plateau effects, which concern the suppression of the underlying gradient magnitude or variance with increasing system size or circuit depth~\cite{McClean2018}. Systematic studies of both gradient estimation costs and trainability limitations will therefore be important for assessing the scalability of the proposed methods.}
Some possible approaches for mitigating these issues include developing a booster method for Meta and NN Meta-VQT similar to the ones designed for VQE by Wang
et al.\cite{booster_vqe}, designing the ansatz with barren plateau free PQC's as mentioned in Deshpande et al.\cite{baren_plateau_free_dynamic_pqc}, incorporating multi-qubit non-unitary operations devised by Za-
pusek et al.\cite{multiqubit_nonunitary} as well as utilizing training schemes that reduces resource costs on quantum hardware as introduced by Bhowmick et al.\cite{bhowmick2025enhancing}.
The neural network chosen for NN Meta-VQT also provides a channel for the scope of improvement when dealing with more complex and larger size Hamiltonians.

Another fundamental challenge, involves the theoretical hardness of Gibbs state preparation itself. For general Hamiltonians, preparing the Gibbs state is a QMA-hard problem. Thus, identifying classes of Hamiltonians for which efficient variational Gibbs state preparation is feasible remains an open theoretical and practical challenge.
Relatedly, for applications like QBMs,  the problem of selecting a suitable Hamiltonian that can faithfully and efficiently capture the underlying data distribution is still not well understood and continues to be an active area of research. In addition, recent works by Mark Wilde \cite{wilde20251,wilde20252} and Adit Vishnu et.al \cite{vishnu2025} proposes different training techniques for improved and scaled QBM implementations. Incorporating our meta-algorithms into these emerging QBM training workflows could therefore represent an interesting direction for future work.

In this work, our comparative study for Gibbs state preparation has been done with only few existing variational technique in literature for specific depth and $\beta$ values.
Though the numerical experiments  suggest that the performance of Meta VQT and NN Meta-VQT is better than that of VarQITE under the specific conditions chosen for our numerical study, there is still scope for comparison with other methods.

The extension of these studies in the continuous variable domain is a possible future direction to explore.
A further important consideration is conducting
error analysis and resource estimation for scaled versions
of these models, particularly in the context of the noisy
hardware on which they are trained or implemented. A proof-of-concept small scale noise analysis of meta-algorithms for Gibbs state preparation of TFIM and Kitaev ring model is presented in Appendix~\ref{noiseanalysis}. Additionally,
a comprehensive understanding of the boundaries
of quantum utility or advantage for such heuristic
methods remains an open question, requiring further theoretical
advancements.

\section{Acknowledgement}
The authors thank Dr.~Quoc Hoan Tran from the Quantum Laboratory at Fujitsu Research Japan, Dr.~Bibhas Adhikari and Dr.~Hannes Leipold from Quantum Laboratory at Fujitsu Research America for their valuable feedback at different stages of the work and for taking the time to review the current draft. The authors extend their immense gratitude to Yasuhiro Endo, Hirotaka Oshima and Shintaro Sato  as well as the entire Robust Quantum Computing Department at Fujitsu Limited for their strategic and technical support.
The code used in this paper will be made publicly available at a later date. A preliminary version of this work was presented at the Fujitsu-IISc Quantum Workshop held in January 2025 in Bangalore, India. Part of this work were also presented as a poster at TQC 2025 and as an invited talk at the First International Workshop on Quantum Boltzmann Machines (IW-QBM), held in December 2025.
\bibliography{ref}
\clearpage
\appendix
\section{System Sizes Demonstrated in Variational Gibbs State Preparation} \label{table_comparison_exisitng_works}
Table~\ref{gibbs_state_exisiting_lit} summarizes the system and ancilla qubit counts reported in representative variational Gibbs state preparation studies, highlighting the range of system sizes demonstrated across existing approaches.
\begin{table*}[t]
\renewcommand{\arraystretch}{2}
\caption{System and ancilla qubit counts in prior variational approaches to Gibbs state preparation.}
\label{gibbs_state_exisiting_lit}
\begin{tabular}{|c|c|c|}
\hline
\multicolumn{1}{|c|}{\textbf{Prior Variational Approaches}} & \multicolumn{1}{|c|}{\textbf{\begin{tabular}[c]{@{}c@{}}System \\ Qubits\end{tabular}}} & \multicolumn{1}{|c|}{\textbf{\begin{tabular}[c]{@{}c@{}}Ancilla \\ Qubits\end{tabular}}} \\
\hline

\makecell[c]{Quantum Hamiltonian-Based Models and the Variational Quantum Thermalizer Algorithm, \\ Verdon et al. \cite{verdon_vqt}} & $4$ & 0 \\

\makecell[c]{Variational quantum Gibbs state preparation with a truncated Taylor series, \\ Wang et al.\cite{wang2005variational}} & 11 & 1 \\

\makecell[c]{Extending the Variational Quantum Eigensolver to Finite Temperatures, \\ Selisko et al.\cite{HVA2}} & 4 & 0 \\

\makecell[c]{Variational quantum simulation of the quantum critical regime, \\ Shi et al.  \cite{simulation_quantum_critical}} & 6 & 0 \\

\makecell[c]{Training Quantum Boltzmann Machines with the $\beta$ Variational Quantum Eigensolver, \\ Huijgen et al.\cite{Huijgen_2024}} & 10 & 0 \\

\makecell[c]{Thermal state preparation of the SYK model using a variational quantum algorithm, \\ Araz et al.\cite{HVA3}} & 12 & 0 \\

\makecell[c]{Dynamic parameterized quantum circuits: expressive and barren-plateau free, \\ Deshpande et al.\cite{baren_plateau_free_dynamic_pqc}} & 10 & 0 \\

\makecell[c]{Dissipative variational quantum algorithms for Gibbs state preparation,\\ Yigal Ilin et al. \cite{dissipative_algos}} & 6 & $0$ \\

\makecell[c]{TEPID-ADAPT: Adaptive variational method for simultaneous preparation \\ of low-temperature Gibbs and low-lying eigenstates, Bharath Sambasivam et al. \cite{sambasivam2025tepidadaptadaptivevariationalmethod}} & 15 & $\approx 2$ \\

\hline
\makecell[c]{\textbf{Meta  and Neural Network Meta-Variational Quantum Thermalizer}, \\ \textbf{Our Work}} & \textbf{8 }& \textbf{8} \\

\hline
\end{tabular}
\end{table*}

\section{Variational Quantum Imaginary Time Evolution (VarQITE)} \label{varqite}
In this method, the approximate Gibbs state is evaluated using PQCs that implements variational quantum imaginary time evolution based on McLachlan's variational principle \cite{mcardle2019variational,Zoufal_circuit_training}.
A brief explanation of how VarQITE is used to approximate the Gibbs state of the Hamiltonian is presented below.

For an initial state, $\ket{\psi_{0}}$, the imaginary time $(\tau = it)$ evolution (ITE) with respect to the Hamiltonian, $H(\vec{h})$, non-unitarily maps the state to,
\eq{ \label{ITE_state_evolution}
\ket{\psi_{\tau}} = \frac{e^{-H(\vec{h}) \tau} \ket{\psi_{0}}}{\sqrt{\bra{\psi_{0}} e^{-2 H(\vec{h}) \tau}\ket{\psi_{0}}}},
}
and follows Wick-rotated Schrodinger's equation,
\eq{\nonumber
\frac{d \ket{\psi_{\tau}}}{d\tau} = -(H(\vec{h}) - E_{\tau})\ket{\psi_{\tau}},
}
where $E_{\tau} = \bra{\psi_{\tau}} H(\vec{h}) \ket{\psi_{\tau}}$ is the system's energy.
If the initial state $\ket{\psi_{0}}$ is taken to be a purification of the maximally mixed state, $\rho = \text{Tr}_{A}\ket{\psi_{0}}\bra{\psi_{0}} = \mathbb{I}/d$, where A represents the ancilla subsystem, and the imaginary time evolution is kept as $\tau \rightarrow 1/2 T = \beta/2$ for the system, then Eq.~\eqref{ITE_state_evolution} yields a purification of the Gibbs state defined in Eq.~\eqref{Gibbs state}.

The ITE state cannot be directly implemented on a quantum circuit because of the non-unitary nature of evolution.
However, McArdle et al.\cite{mcardle2019variational} showed that the ITE can be implemented on a hybrid gate-based quantum computer using McLachlan's variational principle.
The main idea here is as follows. First, the state $\ket{\psi_{\tau}}$ is approximated by defining a parametrized trial state : $\ket{\phi(\Vec{\omega}(\tau))}= U(\Vec{\omega}(\tau))\ket{0}$, where $U(\Vec{\omega}(\tau)) = U_{N}(\omega_{n}) U_{N-1}(\omega_{N-1}).....U_{k}(\omega_{k})....U_{1}(\omega_{1})$ are the set of parametrized single or two-qubit gates with parameters $\vec{\omega}(\tau)$.
Since the goal is the preparation of Gibbs state and this requires purification of maximally mixed states as the initial boundary condition, the state $\ket{\phi(\Vec{\omega}(\tau))}$ is defined over a PQC with a total of $\text{n} + \text{n}_{\text{ancilla}}$ qubits.
The initial value of parameters of the unitaries are chosen such that,
\eq{\nonumber
\ket{\phi(\vec{\omega}(\tau=0))} = U(\vec{\omega}(\tau=0))\ket{0}^{{\otimes}2n} = \ket{\phi^{+}}^{\otimes n},
}
where $\ket{\phi^{+}}^{\otimes n} = \frac{1}{\sqrt{2}} (\ket{00} + \ket{11})$ is the Bell state.
Here, in each Bell state $\ket{\phi^{+}}$, the first qubit belongs to the system, while the second qubit serves as an ancilla.
Tracing out the ancillas from $\ket{\phi(\omega(0))}$ results in the maximally mixed state; $\frac{\mathbb{I}}{2^{n}}$.
Further, the ITE of $\ket{\phi(\Vec{\omega}(\tau))}$ is done by implementing McLachlan variational principle,
\eq{\nonumber
\delta || (\frac{\partial}{\partial \tau} + H(\vec{h}) - E_{\tau}) \ket{\phi({\vec{\omega}(\tau)})} || = 0,}
 using quantum circuits for $\tau = \beta/2$.
The state, $\text{Tr}_{b}[\ket{\phi(\vec{\omega}(\tau=\beta/2))}]$ at the end of $\tau = \beta/2$ evolution gives the approximation to the Gibbs state of our system Hamiltonian, $H(\vec{h})$.
For more detailed calculations, please refer to  Zoufal et al. \cite{Zoufal_circuit_training} and McArdle et al. \cite{mcardle2019variational}.
In this work, the technique has been adapted solely for comparative study using specific examples, with the variational quantum imaginary time evolution simulated using the built-in library available in Qiskit.
\section{Challenges in Helmholtz Free Energy Evaluation and Approximation Approaches} \label{Gibbs free energy evaluation}
Evaluating the exact Helmholtz free energy becomes highly challenging for large system sizes.
 The main difficulty arises from the entropy contribution in Eq.~\ref{Gibbs free energy}, whose computation is generally intractable even on quantum computers for arbitrary large quantum states\cite{entropy_challenge1,entropy_challenge2}.
To address this, several works have proposed algorithms that approximate the Helmholtz free energy while maintaining scalability on near-term devices. Few notable approaches are listed below.
\subsection{Variational Approaches Involving Quantum Circuits}
\begin{itemize}
    \item The work by Wang et al.\cite{wang2005variational} expands the entropy in a Taylor series and truncates it at some finite order
$K=2$ or $K=3$. The truncated series of entropy corresponds to evaluating higher-order overlaps between quantum states. Consequently, Helmholtz free energy can be estimated as a linear combination of the system’s energy and these state overlaps, which are efficiently measurable on near-term quantum hardware. The trade-off is that evaluating higher-order state overlaps requires additional ancilla qubits.
    \item Johannes Selisko et al.\cite{HVA2} introduced the quantum VQT (qVQT) algorithm, which employs two variational quantum circuits (VQCs) together with an intermediate measurement to prepare a mixed state that approximates the Gibbs state. The first VQC produces a probability distribution via intermediate measurement, enabling entropy estimation, while the second VQC evaluates the system’s energy. A classical optimizer iteratively tunes both parameter sets to minimize the Helmholtz free energy. This hybrid quantum-classical scheme provides a resource-efficient and scalable route to Gibbs state preparation, adaptable to different ansatz families. Related approaches have also been applied by Verdon et al. \cite{verdon_vqt} and J. Y. Araz \cite{HVA3} to prepare Gibbs states for models such as the 2D Heisenberg Hamiltonian and the Sachdev–Ye–Kitaev (SYK) model.
    \item Another related approach is the TEPID-ADAPT algorithm \cite{sambasivam2025tepidadaptadaptivevariationalmethod}, which formulates low-temperature Gibbs state preparation through direct minimization of the Helmholtz free energy functional. In this framework, the thermal state is approximated within a truncated variational basis consisting of a small set of low-energy states generated adaptively during the optimization. The mixed state is parametrized as a diagonal density matrix over this truncated basis, allowing the entropy contribution to be computed analytically from the classical probability parameters associated with each basis state, thereby avoiding quantum estimation of the von Neumann entropy. The energy expectation value is evaluated variationally on the quantum device, and the free energy is minimized through a hybrid quantum-classical optimization loop. An alternative ancilla-free implementation estimates the free energy by interpreting the mixed state as a convex combination of pure states from the truncated basis and computing ensemble averages over sampled states.
    \end{itemize}
    \subsection{Classical Tensor Network Optimization followed by Quantum Execution}
    \begin{itemize}
    \item In Ref.~\cite{li2025variationalthermalstatepreparation}, the computational challenge of Helmholtz free energy evaluation is addressed by classically approximating the purified Gibbs state using matrix product states (MPS). The mixed state $\rho$ is embedded into a pure state $|\psi\rangle$ on an enlarged Hilbert space, and the ancilla degrees of freedom are traced out to recover $\rho$. The MPS framework provides an efficient low-rank representation with parameters scaling linearly in system size. After tracing out the ancillas, $\rho$ is expressed as a matrix product operator (MPO), enabling efficient contraction with the Hamiltonian (also in MPO form) to evaluate the energy, while the von Neumann entropy is computed from the Schmidt coefficients obtained via singular value decomposition. Importantly, the entire training process is classical in this method, and only the optimized circuit is run on actual quantum hardware with 36 physical and 8 ancilla qubits to estimate properties of the Gibbs state.
    \end{itemize}
    \subsection{Classical Shadow Approach}
    \begin{itemize}
    \item \rev{Classical shadow methods offer another route to estimating selected thermal state observables from randomized measurements~\cite{huang2020predicting}. In particular, pure thermal shadows have been proposed for predicting linear Gibbs state expectation values, including energy and other observables~\cite{coopmans2023predicting}. Nonlinear properties are not categorically inaccessible: dedicated randomized measurement and shadow-based protocols can estimate quantities such as state moments and $\operatorname{Tr}(O\rho^{2})$~\cite{vermersch2024enhanced, peng2025experimental,du2026optimal}. However, the guarantees for linear expectation values do not directly imply an efficient estimator of the von Neumann entropy of a general trial state. Using shadows for the full free energy objective therefore requires an appropriate entropy estimator, any assumptions it needs, and an analysis of approximation error and sample complexity ~\cite{vermersch2024enhanced,song2026batched}.}
\end{itemize}
\subsection{Markov Entropy Decomposition (MED) as Free Energy surrogate}
\begin{itemize}
    \item  \rev{Poulin and Hastings~\cite{poulin2011markov} introduced a convex relaxation of the finite-temperature free energy problem based on strong subadditivity of von Neumann entropy and locally consistent reduced density operators. They replace the global entropy with an upper bound expressed as a sum of local conditional entropies and relax global consistency to local consistency, obtaining a lower bound on the free energy. The dual optimization problem yields quantum belief-propagation equations, offering a route to finite-temperature calculations without explicitly constructing the full Gibbs density matrix.}

    \item \rev{Scalet et al.~\cite{scalet2025classicalestimationfreeenergy} revisited the Markov entropy decomposition and established convergence guarantees when the effective interaction decays sufficiently rapidly. They obtain polynomial-time approximation algorithms in one dimension at any fixed temperature and quasi-polynomial-time algorithms in a high-temperature regime satisfying an additional commutativity condition. They also show that decay of the effective interaction implies decay of conditional mutual information, and use approximate local marginals to construct a rounding procedure that prepares a global state close to the Gibbs state. Under the stated assumptions, this procedure admits an efficient quantum implementation for approximate Gibbs sampling.}
\end{itemize}

\rev{These approaches illustrate the range of strategies available for handling the entropy contribution, while our focus is on learning a parameter-conditioned state preparation strategy without requiring exact Gibbs states, density matrices, or thermodynamic observables as training labels.}

\section{Overview of Classical and Quantum Boltzmann Machines} \label{overview_qbm}
This section provides brief overview of classical and quantum Boltzmann machines, both of which fundamentally rely on the system's Gibbs state.
\subsection{Classical Boltzmann Machine}
\begin{figure}[!tbh]
    \centering
    \includegraphics[scale=0.20]{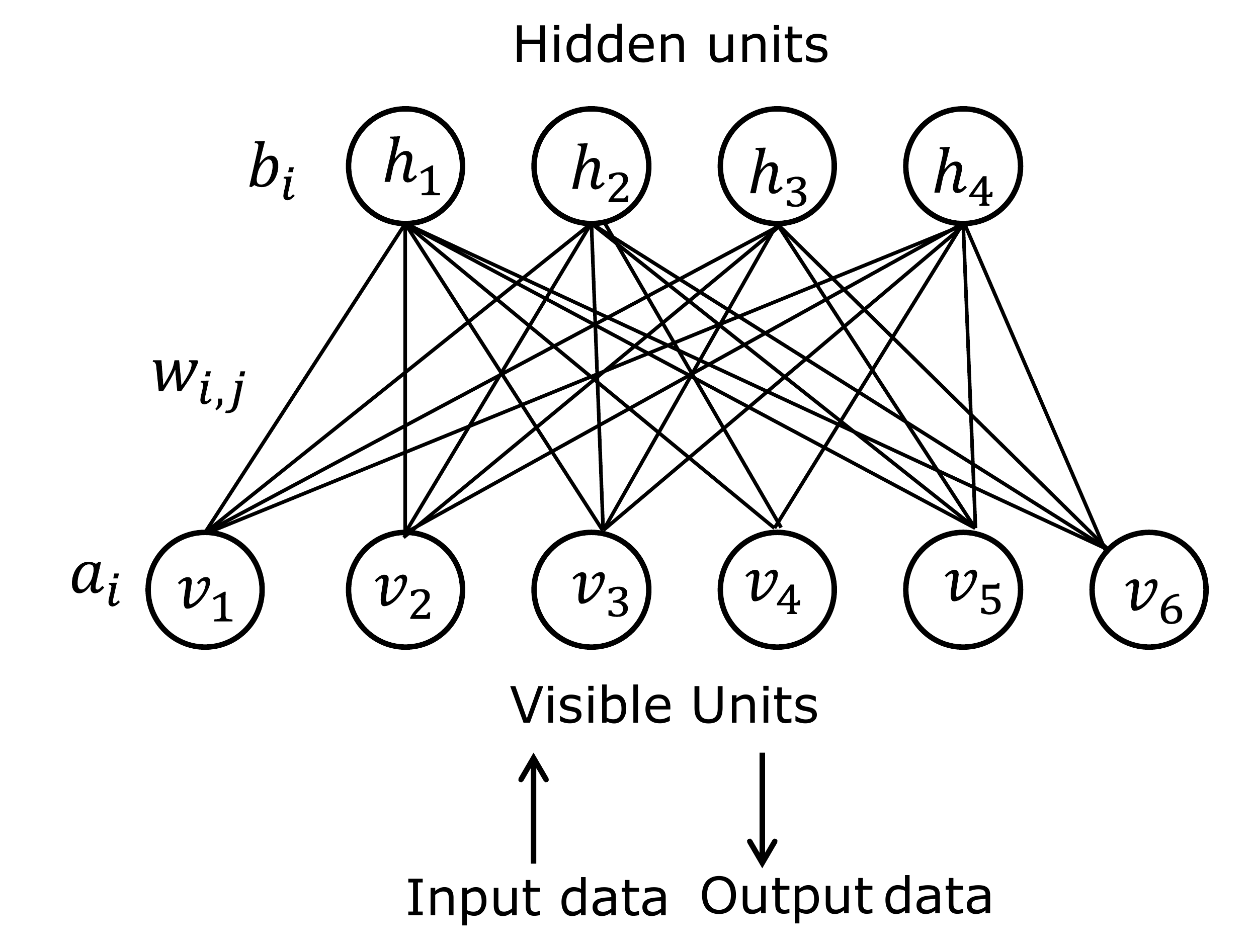}
    \caption{\textbf{Restricted Boltzmann Machine :} The figure illustrates the architecture of a classical restricted Boltzmann machine, where connections exist between visible and hidden neurons but not among neurons within the same layer. $a_{i}$ and $b_{i}$ are the biases of visible and hidden neurons, and $w_{i,j}$ are the connection weights between hidden and visible neurons. The input data is mapped to the visible neurons. The hidden neurons learn the latent features in the input data by adjusting the connection weights and biases. The output is then obtained by sampling the visible layer after a round-trip of Gibbs sampling, where information is propagated from the visible to hidden layer and then back to the visible layer. This reconstructed visible state serves as the model's output, enabling it to generate samples, reconstruct inputs, or estimate probability distributions over input data. }
    \label{rbm}
\end{figure}

Classical Boltzmann Machines are a type of energy-based stochastic neural network inspired by statistical mechanics.
They learn complex probability distributions over input data.
The network consists of interconnected nodes of visible and hidden neurons, that commonly represent binary states.
The input data is represented using the visible neurons while the latent features in the input data is learned by the hidden neurons.
Figure~\ref{rbm}, shows a type of Boltzmann machine known as Restricted Boltzmann Machine (RBM) where connections exist between visible and hidden neurons but not among neurons within the same layer.
The entire network is described by an energy function, $E(v,h) = -\sum_{i=1}^{N_{v}} a_{i}v_{i} - \sum_{i=1}^{N_{h}} b_{i} h_{i} - \sum_{i,j} w_{i,j} v_{i} h_{j}$, where $v$ is the visible unit, $h$ is the hidden unit, $N_{v}$ represents the number of visible units, $N_{h}$ represents the number of hidden units, $a_{i}$ and $b_{i}$ are the visible and hidden unit biases, and $w_{i,j}$ is the connection weight between visible and hidden neurons.
The probability distribution over possible unit configurations is represented by the Boltzmann distribution, $P(v,h) = e^{-\beta E(v,h)}/\sum_{v,h} e^{-\beta E(v,h)}$, which characterizes the state of the system at finite temperature, $T$.
During training, the temperature controls the level of stochasticity in the model, with higher temperatures leading to more uniform sampling across configurations, while lower temperatures bias the system toward lower energy (more probable) states.
The weights connecting the visible and hidden neurons and the biases are trained so that the probability distribution of the target input data is accurately represented by $P(v) = \sum_{h} e^{-\beta E(v,h)}/\sum_{v,h} e^{-\beta E(v,h)}$.
Boltzmann machines are used in various applications, including dimensionality reduction \cite{BM_dimesnionality_reduction}, feature learning \cite{BM_feature_extraction}, collaborative filtering \cite{BM_collaborative_filtering}, and pretraining of deep learning networks \cite{Bm_deep_learning_networks}.
\subsection{Quantum Boltzmann Machine (QBM)}
The quantum version of this classical architecture is known as the Quantum Boltzmann Machine.
The energy function representing the network structure of classical BM shown in Fig.~\ref{rbm}, is replaced by the parametrized Hamiltonian,
\eq{\label{qbm_generic_hamiltonian}
H(\Vec{h}) = \sum_{i=0}^{p-1} h_{i} \sigma_{i},}
where visible and hidden units are now qubits defined by the Pauli matrices $\sigma^{x}, \sigma^{y}$ and $\sigma^{z}$  , $\Vec{h} = (h^{1},h^{2},...h^{p})\, \epsilon \, \mathbb{R}^{p}$ are the trainable coefficients of the Hamiltonian and $\sigma_{i} = \otimes_{j=0}^{n-1} \sigma_{i}^{j}$ with $\sigma$'s as the Pauli matrices and $\sigma_{i}^{j}\, \epsilon \, {I,X,Y,Z}$.
The visible qubits are those that are used to model the input data and measure the model output, while the hidden qubits are those that represent the latent features.
In analogy with classical BM, the aim of QBM is to learn the parameters, $\Vec{h}$, such that the probability distribution of the Gibbs state of the Hamiltonian $H(\Vec{h})$,
approximates the target distributions.
With the inclusion of non-commuting terms in the Hamiltonian which are absent in classical Boltzmann machines, QBMs have the potential for greater expressive power, enabling them to model a broader class of complex data distributions.

For a configuration $\ket{v}$ of the visible qubits, a projective measurement on this visible state is represented as $\hat{\Pi}_{\mathbf{v}} = \ket{v}\bra{v}^{\mathbf{v}} \otimes \mathbb{I}^{\mathbf{h}}$, where $\mathbf{v}$ and $\mathbf{h}$ indicate the visible and hidden subsystems, respectively. The marginal probability of observing the visible configuration $\mathbf{v}$ is then given by,
\eq{\nonumber
p_{\mathbf{v}}(\Vec{h}) = \text{Tr}[\hat{\Pi}_{\mathbf{v}}\rho(\Vec{h})],
}
where $\rho(\vec{h}) = e^{-\beta \hat{H}(\vec{h})}/\text{Tr}[e^{-\beta \hat{H}(\vec{h})}]$ is the Gibbs state of the full Hamiltonian which includes both the visible and hidden qubits.
If $p^{\text{data}}$ is the probability distribution that represents the input data, then the training of QBM involves minimizing the distance between $p^{\text{data}}$ and $p_\mathbf{v}(\Vec{h})$.
The distance between two probability distributions is typically quantified by the  Kullback-Leibler (KL) divergence,
\eq{\nonumber
\mathcal{L}(\Vec{h}) = \sum_{\mathbf{v}} p^{\text{data}} \log \frac{p^{\text{data}}} {p_{\mathbf{v}}(\Vec{h})}.}
The KL divergence always takes a nonnegative value, $L(\vec{h}) \geq 0$, with $L(\vec{h}) = 0$ when $p^{\text{data}} = p_{\mathbf{v}}(\Vec{h})$.
When the input data is provided as an encoded quantum state $\rho^{data}$, the relative quantum entropy $S(\rho^{\text{data}} || \rho(\vec{h})) = \text{Tr}[\rho^{\text{data}}\log \rho^{\text{data}} -\rho^{\text{data}}\log \rho(\vec{h})]$, is used to quantify the difference between the QBM-modeled data and the input target data \cite{bound_based_qbm1}.
Minimization of the KL divergence or relative entropy is done on a classical \cite{Huijgen_2024} or quantum hardware \cite{Zoufal_circuit_training} by evaluating their respective first-order derivatives with respect to coefficient $\vec{h}$ of the Hamiltonian \cite{amin_qbm}.
Furthermore, the optimization is carried out by updating the Hamiltonian parameters based on these gradients as follows,
\eq{\nonumber
\Vec{h}' = \Vec{h} + \eta \partial_{\Vec{h}}\mathcal{L}(\Vec{h}),
}
where $\eta$ is the learning rate.

In this paper, we consider training the QBM with only visible units using our proposed meta-algorithms for Gibbs state preparation. The computation of the gradients of the loss function is done using numerical differentiation.

\begin{table*}[t]
\renewcommand{\arraystretch}{2}
\caption{ \textbf{Robustness to Hamiltonian complexity :} We compare the quality of Gibbs states prepared by NN Meta-VQT and VarQITE as a function of the complexity of a $2$-qubit Hamiltonian.
The precision of NN Meta-VQT, measured by the avearge trace distance between the prepared and true Gibbs state over a range of $(J,h)$ values, remains largely unaffected as the number of non-commuting terms increases, demonstrating robustness to Hamiltonian complexity. In contrast, the accuracy of VarQITE degrades noticeably when the Hamiltonian contains more than two non-commuting operator blocks.}\label{nn_meta_varqite_commuting_blocks}
\begin{tabular}{|c|c|c|c|}
\hline
\textbf{Hamiltonian} & \textbf{Commuting Blocks} & \textbf{NN Meta-VQT} & \textbf{VarQITE} \\ \hline
$H = -J\sigma_{1}^{x}\sigma_{2}^{x}$     & $1$      & $\mathbf{0.04\pm 0.01}$   & $0.002 \pm 0.001$     \\
$H = -J\sigma_{1}^{x}\sigma_{2}^{x}-h\sum_{i=1}^{2} \sigma_{i}^{x}$      & 1      & $\mathbf{0.05 \pm 0.02}$      & $0.005 \pm 0.003$   \\
$H = -J \sigma_{1}^{x}\sigma_{2}^{x} - h \sum_{i=1}^{2} (\sigma_{i}^{x} + \sigma_{i}^{z})$     & 2     & $\mathbf{0.06 \pm 0.02}$      &   $0.016 \pm 0.006$    \\
$H = -J \sigma_{1}^{x}\sigma_{2}^{x} - h \sum_{i=1}^{2} (\sigma_{i}^{x} + \sigma_{i}^{z} + \sigma_{i}^{y})$      &  3      & $\mathbf{0.07 \pm 0.05}$     & $0.33 \pm 0.20$      \\
$H = -J (\sigma_{1}^{x}\sigma_{2}^{x} + \sigma_{1}^{y}\sigma_{2}^{y}) - h \sum_{i=1}^{2} (\sigma_{i}^{x} + \sigma_{i}^{z} + \sigma_{i}^{y})$     & 3     & $\mathbf{0.08 \pm 0.05}$    & $0.30 \pm 0.17$      \\
$H = -J (\sigma_{1}^{x}\sigma_{2}^{x} + \sigma_{1}^{y}\sigma_{2}^{y} + \sigma_{1}^{z}\sigma_{2}^{z}) - h \sum_{i=1}^{2} (\sigma_{i}^{x} + \sigma_{i}^{z} + \sigma_{i}^{y})$   & 3      &  $\mathbf{0.08 \pm 0.03}$      &  $0.38 \pm 0.17$      \\ \hline
\end{tabular}
\end{table*}
\begin{table}[h!]
\renewcommand{\arraystretch}{1.5}
\caption{\textbf{ Comparsion between Meta-VQT, NN Meta-VQT and VarQITE :} For the 2-qubit Heisenberg Hamiltonian with external fields shown in Eq.~\eqref{complex_hamiltonian}, the Meta-VQT and NN Meta-VQT algorithms prepare the Gibbs state with much better precision than VarQITE technique with similar circuit depth.}\label{table_meta_nn_varqite}
\begin{tabular}{|c|c|c|}
\hline
\textbf{Method} & \textbf{Layers} & \textbf{Average Trace distance} \\
\hline
Meta-VQT & 4 SU2 + 1 HVA & $0.18 \pm 0.07$ \\
Meta-VQT & 4 SU2 & $0.22 \pm 0.09$ \\
NN Meta-VQT & 4 SU2 + 1 HVA & $0.15 \pm 0.02$ \\
NN Meta-VQT & 4SU2 & $0.08 \pm 0.03$ \\
VarQITE & 4SU2 & $0.32 \pm 0.20$ \\
\hline
\end{tabular}
\end{table}

\section{Classical optimization overhead for meta-algorithms}\label{Classical_optimization_overhead}

For Meta-VQT across all system sizes, we employed the Adam optimizer to update both the encoding (HEA) and processing ansatz (HVA) parameters, with an initial learning rate of $0.1$. For larger system sizes ($N \geq 4$), convergence of the loss function typically requires up to $1000$ optimization steps, whereas for smaller systems convergence is achieved within approximately $600$ iterations.

The approximate wall-clock times required for convergence for system sizes $N = 2, 3, 4, 6,$ and $8$ are respectively
$6~\text{min},\; 11~\text{min},\; 30~\text{min},\; 110~\text{min},\; \text{and } 38~\text{hours}$.

For the NN Meta-VQT training, the Adam optimizer was used to update all classical neural network parameters. The initial learning rate was set to $10^{-2}$ for system sizes $N = 2$ and $3$, $10^{-3}$ for $N = 4$, and $10^{-4}$ for $N = 6$ and $8$. To gradually reduce the learning rate during training, we employed a StepLR scheduler with decay factor
\[
\gamma = \exp\left(\frac{\log(0.7)}{1000}\right),
\]
such that after $1000$ optimization steps the learning rate is reduced to $70\%$ of its initial value. The scheduler updates the learning rate at every optimization step (step\_size = 1), resulting in an effective exponential decay profile throughout training.

The total number of optimization steps required for convergence is approximately $1000$ for system sizes $N \leq 4$ and about $1200$ for $N \geq 6$. The corresponding wall-clock times for NN Meta-VQT for $N = 2, 3, 4, 6,$ and $8$ are
$11~\text{min},\; 18~\text{min},\; 31~\text{min},\; 110~\text{min},\; \text{and } 40~\text{hours}$, respectively.

All reported wall-clock times correspond to simulations of the TFIM using HEA and HVA layer configurations $[n,n]$. The computational cost depends sensitively on the chosen ansatz depth and model configuration. In particular, for NN Meta-VQT, inclusion of HVA layers is not strictly necessary to achieve high accuracy; in such cases, the wall-clock time can be significantly reduced relative to standard Meta-VQT.

All numerical simulations were performed on a CPU-based classical computing system (standard laptop architecture) without GPU acceleration.

\section{Pseudoalgorithms} \label{pseudoalgo}
The pseudoalgorithms to run Meta-VQT, NN Meta-VQT and NN Meta-VQT based QBM are shown in Algorithm~\ref{Meta-VQT pseudo}, Algorithm~\ref{NN Meta-VQT pseudo} and Algorithm~\ref{variational QBM pseudo}, respectively.
\noindent
\begin{algorithm}[H]
\caption{Meta-Variational Quantum Thermalizer}
\label{Meta-VQT pseudo}
\begin{center}\textbf{Input:} Parameter $\Vec{h}$ of Hamiltonian \\
\textbf{Output:} Quantum Gibbs state of $H(\Vec{h})$\end{center}
\begin{algorithmic}
\vspace{0.1cm}
\State // Initialization
\vspace{0.1cm}
\State $\beta \gets 1$ \hfill // Inverse temperature
\State $n_\text{ancilla} \gets n_\text{system}$ \hfill // Number of ancillas
\State $n_\text{proc} \gets n_\text{enco} \gets n_\text{system}$ \hfill // Number of circuit layers
\State $\Vec{h}_\text{train}$ \hfill // Training parameter set
\State $n_\text{epochs}$ \hfill // Number of training steps
\State $\eta \gets 0.01$ \hfill // Learning rate
\State $(\Vec{\omega}, \Vec{\theta}, \Vec{\phi}) \gets (\Vec{\omega}_0, \Vec{\theta}_0, \Vec{\phi}_0)$ \hfill // Initial circuit parameters
\vspace{0.1cm}
\State // Training Loop
\vspace{0.1cm}
\For{$n = 0$ \textbf{to} $n_\text{epochs}$}
    \State Loss $\gets 0$
    \ForAll{$\Vec{h}_i \in \Vec{h}_\text{train}$}
        \State EncoAnsatz $\gets \mathrm{SU2}(\Vec{h}_i, \Vec{\omega}, \Vec{\phi})$
        \State ProcAnsatz $\gets \mathrm{HVA}(\Vec{\theta})$
        \State Ansatz $\gets$ EncoAnsatz + ProcAnsatz
        \State Output state $\gets \text{Simulate}(\text{Ansatz})$
        \State $\tilde{\rho} \gets \text{Trace out ancillas}(\text{Output state})$
        \State $G(\Vec{h}_i) \gets \text{Evaluate Helmholtz free energy}(\tilde{\rho})$
        \State Loss $\gets$ Loss + $G(\Vec{h}_i)$
    \EndFor
    \State Global loss $\gets$ Loss
    \State Gradients $\gets \text{Compute gradients}(\text{Global loss})$
    \State $(\Vec{\omega}, \Vec{\theta}, \Vec{\phi}) \gets (\Vec{\omega}, \Vec{\theta}, \Vec{\phi}) - \eta \cdot \text{Gradients}$
\EndFor

\State \Return $\text{Ansatz}(\Vec{\omega}, \Vec{\theta}, \Vec{\phi})$
\end{algorithmic}
\end{algorithm}
\begin{algorithm}[H]
\caption{Neural Network Meta-Variational Quantum Thermalizer}
\label{NN Meta-VQT pseudo}
\begin{center}\textbf{Input:} Parameter $\vec{h}$ of Hamiltonian \\
\textbf{Output:} Quantum Gibbs state of $H(\vec{h})$\end{center}
\begin{algorithmic}
\State // Initialization
\vspace{0.1cm}
\State $\beta \gets 1$ \hfill // Inverse temperature
\State $n_{\text{ancilla}} \gets n_{\text{system}}$ \hfill // Number of ancillas
\State $n_{\text{proc}} \gets n_{\text{enco}} \gets n_{\text{system}}$ \hfill \ // Number of circuit layers
\State $\vec{h}_{\text{train}}$ \hfill // Training parameter set
\State $n_{\text{epochs}}$ \hfill // Number of training steps
\State $\eta \gets 0.001$ \hfill // Learning rate
\State $\vec{\phi}_{\text{nn}} \gets \vec{\phi}^{0}_{\text{nn}}$ \hfill // Initial neural network parameters
\vspace{0.1cm}
\State // Training Loop
\vspace{0.1cm}
\For{$n = 0$ to $n_{\text{epochs}}$}
  \State $\text{Loss} \gets 0$
  \For{$\vec{h}_{i} \in \vec{h}_{\text{train}}$}
  \State $\text{Encoding neural network} \gets (\vec{h}_{i},\vec{\phi}_{nn})$
    \State $\vec{\theta} \gets \text{Encoding neural network}(\vec{h}_{i}, \vec{\phi}_{\text{nn}})$
    \State $\text{Ansatz} \gets \text{SU2}(\vec{\theta}^{(1)}) + \text{HVA}(\vec{\theta}^{(2)})$
    \State $\text{Output state} \gets \text{Simulate}(\text{Ansatz})$
    \State $\tilde{\rho} \gets \text{Trace out ancillas}(\text{Output state})$
    \State $G(\vec{h}_{i}) \gets \text{Evaluate Helmholtz free energy} (\tilde{\rho})$
    \State $\text{Loss} \gets \text{Loss} + G(\vec{h}_{i})$
  \EndFor
  \State $\text{Global loss} \gets \text{Loss}$
  \State $\text{Gradients} \gets \text{Compute gradients}(\text{Global loss})$
  \State $\vec{\phi}_{\text{nn}} \gets \vec{\phi}_{\text{nn}} - \eta \cdot \text{Gradients}$
\EndFor
\State // Return final circuit
\State \Return $\text{Ansatz}(\vec{\phi}_{\text{nn}})$
\end{algorithmic}
\end{algorithm}
\begin{algorithm}[H]
\caption{Variational QBM with NN Meta-VQT} \label{variational QBM pseudo}
\textbf{Input:} Probability distribution of a random quantum state \\
\textbf{Output:} Hamiltonian whose Gibbs states models the input distribution.
\begin{algorithmic}
\vspace{0.1cm}
\State // Initialization \\
\vspace{0.1cm}
$p_{target}$ \hfill // Choose target probability distribution \\
$H(\Vec{J},\Vec{h})$ \hfill // Choose model Hamiltonian  \\
$(\Vec{J},\Vec{h})$ \hfill // Trainable coefficients of Hamiltonian  \\
 $\text{Ansatz}(\Vec{\phi}_{nn})$ \hfill // Trained NN Meta-VQT ansatz from Algorithm~\ref{NN Meta-VQT pseudo} \\
$n_{epochs}$ \hfill // Number of training steps \\
$\eta \gets 0.1$ \hfill // Learning rate
\vspace{0.1cm}
\State // Training Loop
\vspace{0.1cm}
\For{$n = 0$ \textbf{to} $n_{epochs}$}
    \State Ansatz $(\Vec{\phi}_{nn})$ $\gets$ $(\Vec{h}, \Vec{J})$
    \State Output state  $\gets$ Ansatz $(\Vec{\phi}_{nn}) (\vec{h},\Vec{J})$
    \State $\tilde{\rho} \gets$ Trace out ancilla $(\text{Output State})$
    \State $p_{var} \gets$ Measure diagonal elements$(\tilde{\rho})$
    \State Loss $\gets$ KL divergence $(p_{var}, p_{target})$
   \State Gradients $\gets$ Compute gradients (Loss)
   \State $(\Vec{J},\Vec{h}) \gets (\Vec{J},\Vec{h}) - \eta\cdot \text{Gradients}$.
\EndFor
\State \Return $(\Vec{J},\Vec{h})$
\end{algorithmic}
\end{algorithm}
\section{Purity of TFIM Gibbs state prepared using meta-algorithms}\label{purityplots}
\begin{figure*}[!tbh]
    \centering
    \includegraphics[width=1\textwidth, height = 175pt]{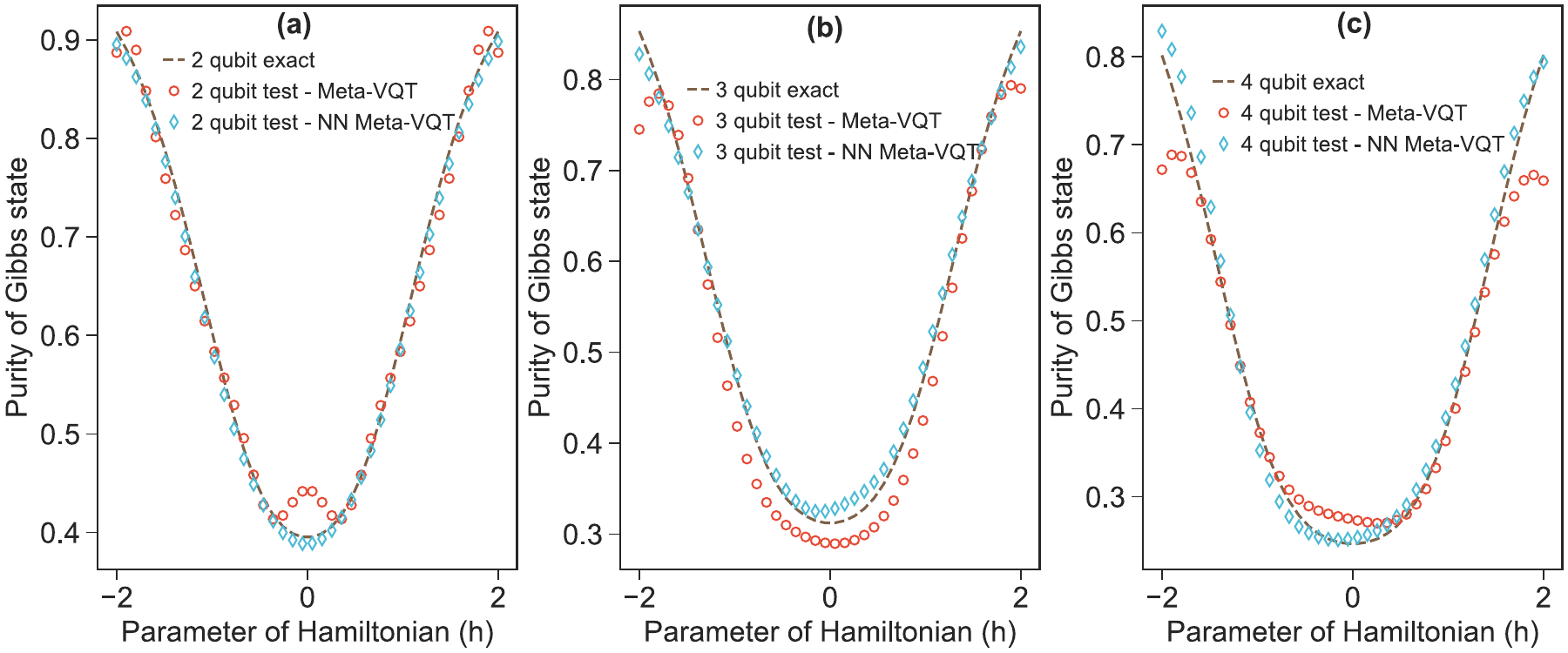}
    \caption{\textbf{Purity of TFIM Gibbs state prepared using Meta-VQT and NN Meta-VQT:} The dashed lines in panels (a), (b), and (c) correspond to the purity of the exact Gibbs state of the TFIM for system sizes $n=2,3~\&~4$, respectively. The meta-algorithms are trained on $10$ uniformly spaced values in the interval $[-2,2]$. The Gibbs state is evaluated on a test set consisting of $40$ uniformly spaced points in $[-2,2]$. The red markers denote purity of the Gibbs state prepared using Meta-VQT, while the cyan markers indicate purity of the Gibbs state prepared using NN Meta-VQT. Both meta-algorithms are able to prepare the Gibbs state with accuracy close to that of the exact state across system sizes $n=2,3~\&~4$.}
    \label{purity_meta_nnmeta}
\end{figure*}

The purity of Gibbs state prepared using Meta-VQT and NN Meta-VQT is plotted and compared with the purity of exact Gibbs state for TFIM of size $n=2,3~\&~4$ in Fig.~\ref{purity_meta_nnmeta} (a), (b) and (c). Both meta-algorithms are able to prepare Gibbs state with purity close to the exact Gibbs state. Similar to our observation from Fig.~\ref{results_tfim_nnmetavqt}, among both meta-algorithms, NN Meta-VQT seems to have slightly better accuracy than the Meta-VQT.
\begin{figure}
    \centering
    \includegraphics[width=1\linewidth]{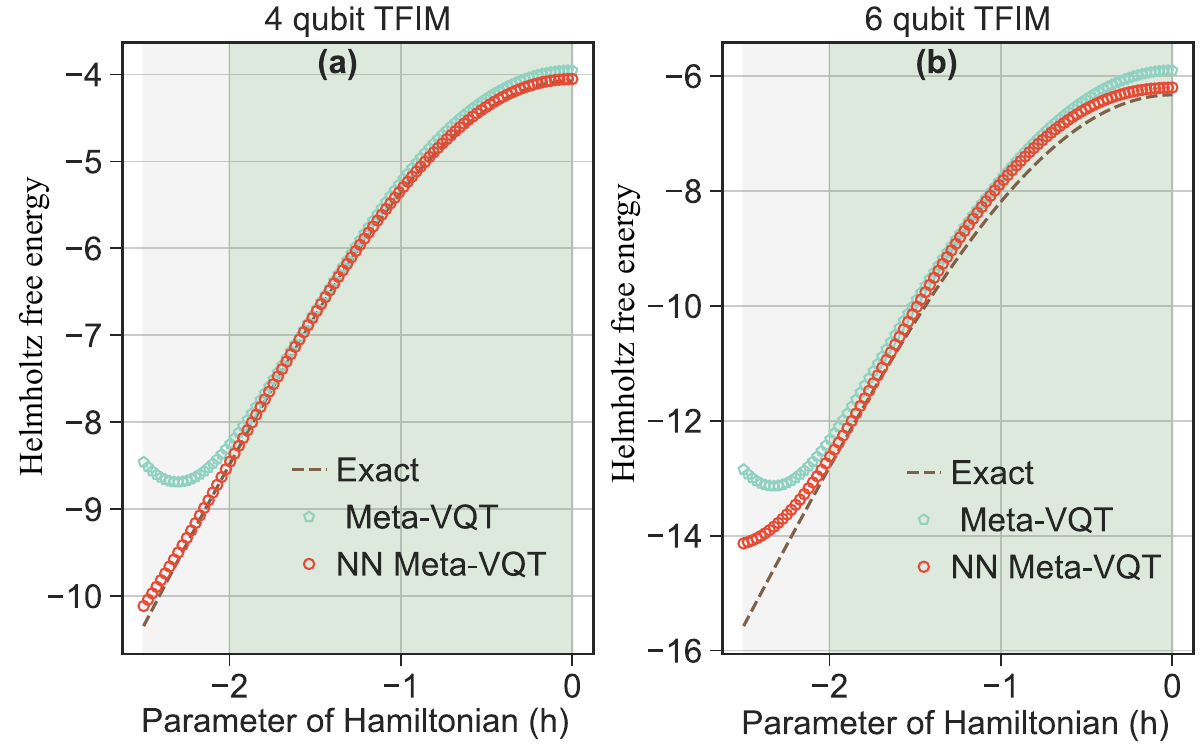}
    \caption{\textbf{Helmholtz free energy profile from meta-algorithms outside the training interval: } In (a) and (b), the dashed lines correspond to the exact Helmholtz free energy evaluated at $100$ uniformly spaced points in $h \in [-2.5,0]$ for a TFIM of size $n=4$ and $n=6$. The meta-algorithms are trained on $5$ uniformly spaced values in the interval $h \in [-2,0]$. The cyan markers in (a) and (b) denote the Helmholtz free energy obtained from the variationally prepared Gibbs state using the trained Meta-VQT algorithm for system sizes $n=4$ and $6$, while the red color markers correspond to the Helmholtz free energy obtained from states prepared using trained NN Meta-VQT algorithm. For $n=4$, the NN Meta-VQT is able to generate the Helmholtz free energy accurately to the exact one across the test set. The accuracy for $n=4$ Meta-VQT and $n=6$ Meta and NN Meta-VQT is observed to decrease outside the training range, particularly for $h \leq -2$.}
    \label{GE_outsidetraining_metvqt}
\end{figure}
\begin{figure}
    \centering
    \includegraphics[width=1\linewidth]{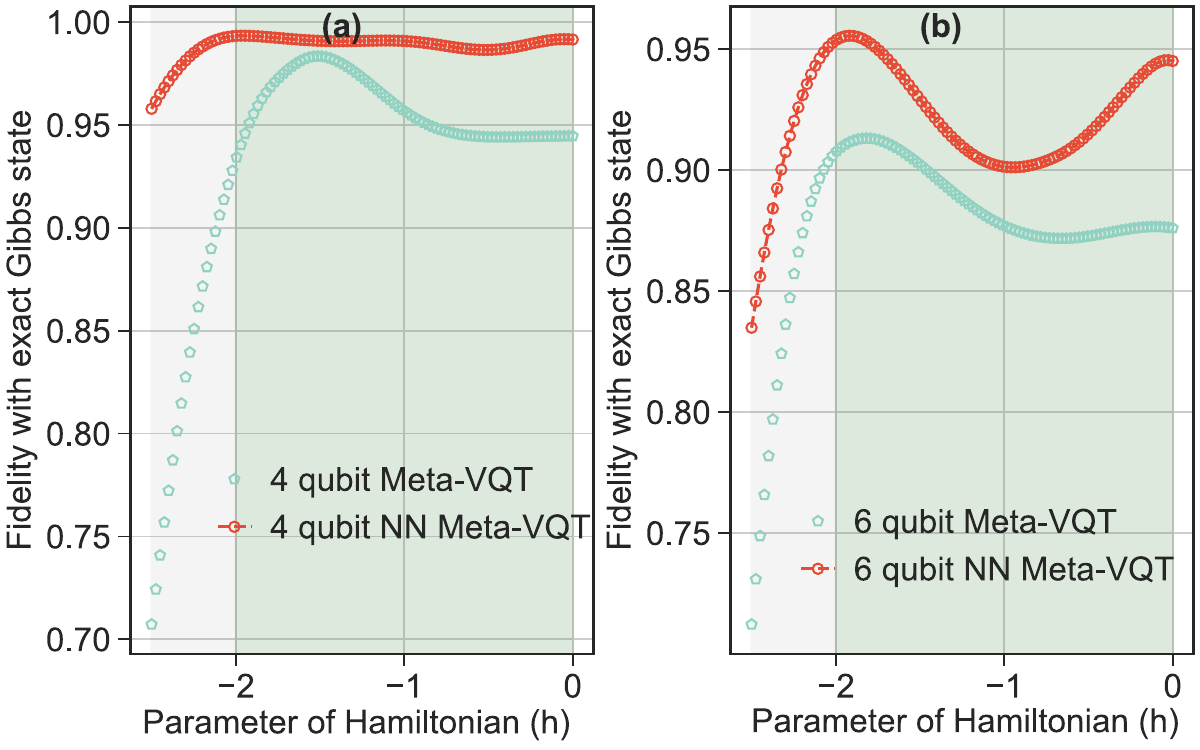}
    \caption{\textbf{Fidelity between exact Gibbs state and the Gibbs state  prepared using meta-algorithms outside the training interval: }  Panels (a) and (b) show the fidelity between the exact and variationally prepared Gibbs states using the trained meta-algorithms, evaluated at $100$ uniformly spaced points in $h \in [-2.5,0]$ for the TFIM with system sizes $n=4$ and $n=6$. The meta-algorithms are trained on $5$ uniformly spaced values in the interval $h \in [-2,0]$. The cyan markers in (a) and (b) correspond to fidelity of states from Meta-VQT algorithm, while the red marker correpsonds to fidelity of states prepared from NN Meta-VQT algorithm. For $n=4$, the fidelity of NN Meta-VQT algorithm is found to be between $0.95$ to $1$. Generally, the accuracy is observed to decrease outside the training range, particularly for $h \leq -2$. However, the degradation is not severe, indicating that even outside the training domain the models predict states reasonably close to those corresponding to unseen points within the training range. }
    \label{fidelity_outside_trainingset}
\end{figure}
\section{Gibbs state preparation outside the range of training set for TFIM}\label{results_outsidetraining}

To assess the extrapolation capability of the proposed meta-algorithms, we analyze their performance for unseen points within and beyond the training interval for the TFIM. The models are trained on $5$ uniformly spaced values in the interval $h \in [-2,0]$ and subsequently evaluated on an extended domain $h \in [-2.5,0]$ using $100$ uniformly spaced test points.

The Helmholtz free energy profiles obtained using Meta-VQT and NN Meta-VQT are shown in Fig.~\ref{GE_outsidetraining_metvqt} for system sizes $n=4$ and $n=6$. For $n=4$, NN Meta-VQT closely reproduces the exact Helmholtz free energy across the test domain, while Meta-VQT exhibits a mild degradation outside the training range. For $n=6$, both Meta-VQT and NN Meta-VQT show a reduction in accuracy as the parameter moves beyond the training interval, particularly for $h \leq -2$.

\rev{These results demonstrate that the learned parameter-conditioned
state preparation strategies can extend beyond the sampled training
domain, with NN Meta-VQT exhibiting particularly strong local
extrapolation for the smaller system studied. The present tests provide
a quantitative characterization of the local extrapolation regime,
while the extent to which this behavior persists across broader parameter ranges, larger system sizes, and different Hamiltonian families requires further study.}
\begin{figure} [!tbh]
    \centering
    \includegraphics[width=1\linewidth]{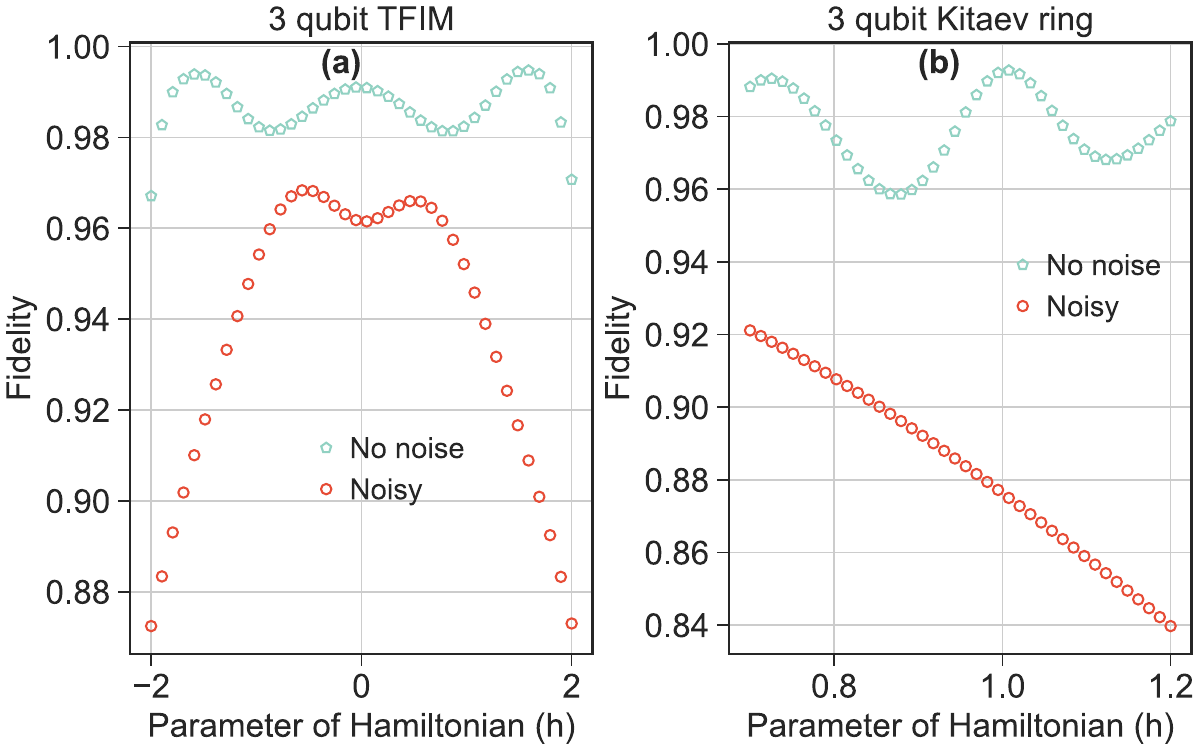}
    \caption{\textbf{Fidelity under noisy simulation : } (a) and (b) show the fidelity between the exact Gibbs state and the state prepared using the trained Meta-VQT for the test points under noiseless (cyan) and noisy (red) simulations for the TFIM and the Kitaev ring model, respectively. The system size considered in both cases is $3$ qubits. The TFIM training set consists of $10$ points sampled from $h \in [-2,2]$, while the Kitaev ring training set contains $20$ points from $h \in [0.9,1.2]$. In both models, $40$ test points are used to generate these plots. The noisy simulations are performed by incorporating a depolarizing channel with error probabilities $0.001$ for single-qubit gates and $0.01$ for two-qubit gates. As expected, the performance degrades in the presence of noise; however, the prepared state still retains the essential properties of the Gibbs state and the fidelity does not drop significantly.}
    \label{noisy_sim}
\end{figure}
\section{Preparing Gibbs state of a Heisenberg Hamiltonian with external field} \label{heisenberg}
To further expand the scope, the Meta and NN Meta- VQT is trained to generate the quantum Gibbs state of a $2$-qubit Heisenberg Hamiltonian with external fields,
\begin{multline}\label{complex_hamiltonian}
 H(J,h) = -J(\sigma_{1}^{x}\sigma_{2}^{x} + \sigma_{1}^{y}\sigma_{2}^{y} + \sigma_{1}^{z}\sigma_{2}^{z})- \\ h(\sum_{i=1}^{2} (\sigma_{i}^{z} + \sigma_{i}^{x} + \sigma_{i}^{y})).
\end{multline}
The parameter values used for training Meta-VQT (NN Meta-VQT) are $10$ uniformly (randomly) selected values between $-2$ and $2$ for each parameter $J$ and $h$.
The average trace distance between the prepared and exact Gibbs state is evaluated and compared with the VarQITE technique existing in literature as shown in Table \ref{table_meta_nn_varqite}.

The quantum circuit depth chosen for VarQITE is $4$ layers of SU2 hardware efficient ansatz.
For Meta-VQT and NN Meta-VQT, two cases are considered, one where the ansatz is a $4$ layered SU2 hardware efficient circuit and the other with $4$ SU2 and $1$ HVA layers.
The number of ancillas required by the two meta-algorithms and VarQITE is same as the system size which is $2$.
From Table \ref{table_meta_nn_varqite}, the average trace distance between the true Gibbs state and the one evaluated from Meta and NN Meta-VQT algorithm for all four cases is lower and more precise than the avearge trace distance evaluated between true Gibbs state and the one prepared using VarQITE.
Additionally, the NN Meta-VQT algorithm generates good quality Gibbs state even in the absence of HVA, while the Meta-VQT with at least one additional layer of HVA can improve the quality of Gibbs state than the one without HVA.
\section{Complexity of the Hamiltonian and the efficiency of NN Meta-VQT}\label{complexity}
In this section, we present results from another numerical study that does an analysis of the dependency of quality of prepared Gibbs state using different algorithms and the complexity of the Hamiltonian.
For the numerics, the NN Meta-VQT with $4$ SU2 hardware efficient layers is considered for an equivalent comparison with the VarQITE that also have $4$ SU2 hardware efficient layers.
The complexity of the Hamiltonian is analyzed using number of commuting blocks, which also indicates the number of non-commuting terms in the Hamiltonian \cite{Hamiltonian_partitioning}.
The average trace distance between the target and prepared Gibbs state from VarQITE and NN Meta-VQT as a function of complexity of Hamiltonian is shown in Table \ref{nn_meta_varqite_commuting_blocks}.
For the tested two-qubit Hamiltonians, the average trace distance of VarQITE increases with the number of commuting blocks, whereas the NN Meta-VQT values remain comparatively stable over the limited range considered.
Additionally, for the tested Hamiltonians with more than $2$ commuting blocks, NN Meta-VQT yields a lower average trace distance than VarQITE.
\section{Noise analysis on small systems}\label{noiseanalysis}
To assess the robustness of our meta-algorithms against quantum hardware noise, we perform a proof-of-concept noise analysis on the three-qubit transverse-field Ising model (TFIM) and the three-qubit Kitaev ring model. Depolarizing noise is incorporated at the circuit level, with error probabilities set to $p_1 = 0.001$ for single-qubit gates and $p_2 = 0.01$ for two-qubit gates, since $2$ qubit gates are generally an order of magnitude more error-prone than single-qubit gates.

For the TFIM, the Meta-VQT model is trained using 10 uniformly spaced values of the transverse field $h \in [-2,2]$. For the Kitaev ring model, 20 uniformly spaced points in the interval $h \in [0.9,1.2]$ are used for training, covering the finite-temperature crossover regime. After training, performance is evaluated on a denser test set consisting of 40 uniformly spaced points within the same parameter intervals.

To quantify robustness, we compute the fidelity between the Gibbs state prepared by the trained Meta-VQT under noisy simulation and the corresponding exact Gibbs state. These fidelities are compared against those obtained in the noiseless setting for the same test points. The results are presented in Fig.~\ref{noisy_sim}(a) and (b). We observe that the introduction of depolarizing noise does not lead to severe performance degradation. While accuracies decrease relative to the noiseless case, the reduction remains moderate and within acceptable bounds for near-term quantum simulations.

We further observe that the fidelity under noise depends on the Hamiltonian parameters. This variation may be related to intrinsic properties of the thermal state, such as purity or entanglement structure, though a systematic investigation of this dependence requires dedicated study.

Although we focus on Meta-VQT in this analysis, we find that the NN Meta-VQT exhibits comparable robustness under identical noise conditions. For the small system sizes considered here, our results demonstrate encouraging resilience of the proposed meta-learning framework to the specific choice of gate errors.

We emphasize, however, that a detailed scalability study under noise remains an open question. A systematic investigation of device noise as well as shot noise effects in larger systems, deeper circuits, hardware executions and the incorporation of error-mitigation protocols is left for future work.
Finally, it is worth noting that Gibbs states, being intrinsically mixed, can potentially exhibit a degree of natural robustness to noise compared to pure states.

\end{document}